\documentclass[12pt]{article}
\usepackage{latexsym}
\usepackage{mathrsfs}
\usepackage{amssymb}
\usepackage{amsmath}
\usepackage{amsfonts, epsf,epsfig,color}
\usepackage{graphicx}
\usepackage{tensor}
\usepackage[all]{xy}

\newcommand{\mcE} {\mathcal{E}}

\newcommand{\mcI} {\mathcal{I}}

\newcommand{\mcO} {\mathcal{O}}

\newcommand{\mcQ} {\mathcal{Q}}

\newcommand{\ii} {\mathrm{i}}

\newcommand{\ind} {\indices}

\newcommand{\lb} [1] {{\left[ #1 \right. }}
\newcommand{\rb} [1] {{\left. #1 \right] }}

\newcommand{\Spin} {\mathrm{Spin}}

\newcommand{\SO} {\mathrm{SO}}

\newcommand{\U} {\mathrm{U}}

\newcommand{\Ss} {\mathbb{S}}

\newcounter{mnotecount}[section]
\renewcommand{\themnotecount}{\thesection.\arabic{mnotecount}}

\newcommand{\mnote}[1]
{\protect{\stepcounter{mnotecount}}$^{\mbox{\footnotesize
$
\bullet$\themnotecount}}$ \marginpar{
\raggedright\tiny\em
$\!\!\!\!\!\!\,\bullet$\themnotecount: #1} }

\newcommand{\C}{\mathbb{C}}

\newcommand{\CP}{\mathbb{CP}}
\newcommand{\RP}{\mathbb{RP}}
\newcommand{\PT}{\mathbb{PT}}
\newcommand{\R}{\mathbb{R}}
\renewcommand{\P}{\mathbb{P}}

\newcommand{\PS}{\mathbb{PS}}
\newcommand{\PPS}{\mathbb{PPS}}

\newcommand{\F}{\mathbb{F}}
\newcommand{\bbH}{\mathbb{H}}

\newcommand{\M}{\mathbb{M}}

\newcommand{\T}{\mathbb{T}}
\newcommand{\Z}{\mathbb{Z}}
\newcommand{\p}{\partial}
\newcommand{\e}{\mathrm{e}}

\newcommand{\D}{\mathrm{D}}

\newcommand{\cO}{\mathcal{O}}

\renewcommand{\P}{\mathbb{P}}
\newcommand{\SL}{\mathrm{SL}}

\newcommand{\SU}{\, \mathrm{SU}}

\newcommand{\rd}{\, \mathrm{d}}

\newcommand{\be}{\begin{equation}\label}
\newcommand{\ee}{\end{equation}}
\newcommand{\bea}{\begin{eqnarray}\label}
\newcommand{\eea}{\end{eqnarray}}

\newcommand{\dbar}{\bar\p}
\newtheorem{defn}{Definition}[section]

\newtheorem{propn}{Proposition}[section]

\begin{document}

\title{\textbf{Conformal Field Theories \\in\\ Six-Dimensional Twistor Space}}

\author{L. J. Mason$^{\star}$, R. A. Reid-Edwards$^{\star}$ and A. Taghavi-Chabert$^{\dagger}$\\
\\
\small{$^{\star}$The Mathematical Institute, 
24-29 St.~Giles, Oxford, OX1 3LB, UK}
\\
\small{$^{\dagger}$Masaryk University, Faculty of Science, Department of Mathematics and Statistics,}\\
\small{
 Kotl\'{a}\v{r}sk\'{a} 2, 611 37 Brno,
Czech Republic}}
\date{}

\maketitle


\begin{abstract}
  This article gives a study of the higher-dimensional Penrose
  transform between conformally invariant massless fields on
  space-time and cohomology classes on twistor space, where twistor
  space is defined to be the space of projective pure spinors of the
  conformal group.  We focus on the 6-dimensional case in which
  twistor space is the six-quadric $Q$ in $\CP^7$ with a view to
  applications to the self-dual $(0,2)$-theory.  We show how
  spinor-helicity momentum eigenstates have canonically defined
  distributional representatives on twistor space (a story that we
  extend to arbitrary dimension).  These give an elementary proof of
  the surjectivity of the Penrose transform.  We give a direct
  construction of the twistor transform between the two different
  representations of massless fields on twistor space ($H^2$ and
  $H^3$) in which the $H^3$s arise as obstructions to extending the
  $H^2$s off $Q$ into $\CP^7$.  

  We also develop the theory of Sparling's `$\Xi$-transform', the
  analogous totally real split signature story based now on real
  integral geometry where cohomology no longer plays a role.  We
  extend Sparling's $\Xi$-transform to all helicities and
  homogeneities on twistor space and show that it maps kernels and
  cokernels of conformally invariant powers of the ultrahyperbolic
  wave operator on twistor space to conformally invariant massless
  fields on space-time.  This is proved by developing the
  6-dimensional analogue of the half-Fourier transform between
  functions on twistor space and momentum space.  We give a
  treatment of the elementary conformally invariant $\Phi^3$ amplitude
  on twistor space and finish with a discussion of
   conformal field theories in twistor
  space.
\end{abstract}

\newpage
\tableofcontents

\section{Introduction}

Twistor methods have become a powerful tool in the study of
four-dimensional gauge theories \cite{Witten:2003nn,Boels:2006ir,
  ArkaniHamed:2009dn,Adamo:2011cb} and the question arises as to how it
might interact  with other important techniques such as AdS/CFT and
reductions of $(2,0)$ self-dual gerbe theory in six dimensions.  This
article is a study of the appropriate higher-dimensional twistor
theory that applies to such higher dimensional theories with a
focus on six dimensions.   Ultimately we would like to
encode interacting field theories on twistor space in higher
dimensions, but this paper will focus mostly on the twistor correspondence for
linear fields (with the exception of a discussion of the $\phi^3$
vertex for scalar fields).

The Penrose transform in higher dimensions and indeed for general
classes of homogeneous spaces was developed in the 1980s with the
general framework summarized in \cite{Baston:1989vh}.  In higher even
dimension $2m$, twistor space can be taken to be the space of totally null self-dual $m$-planes so that twistor space is the projective self-dual
chiral pure spinor space of the conformal group \cite{Penrose:1986ca}.   There is a Penrose transform from conformally invariant massless fields on space-time to
cohomology classes on regions in twistor space.  In six dimensions the
conformal group is a real form of $\SO(8,\C)$ and we have a version of
triality so that the chiral spinor representations for the conformal group are eight-dimensional and are endowed with a quadratic form like the fundamental one.  This is the first
dimension in which the purity condition is nontrivial being the
condition that the chiral spinors should be null with respect to the
quadratic form.  Thus, twistor space turns out to be the six-
dimensional complex quadric $Q$ in $\CP^7$.  As a manifold it is the
same as complexified space-time, but the representation of the conformal group and its correspondence with space-time is quite different.  Conformally invariant massless fields on space-time are represented on $Q$ both as $H^2$ and as $H^3$ cohomology
classes (i.e., closed Dolbeault $(0,2)$ or $(0,3)$-forms modulo exact
ones).  As far as massless fields are concerned, Maxwell and
linearized gravity are no longer conformally invariant, but the wave
equation and symmetric spinor fields satisfying higher spin versions
of the massless Dirac equation are, and it is these fields of a fixed
chirality that are most straightforwardly represented on twistor
space.  The abelian version of the $(2,0)$ self-dual gerbe theory can
be built from such ingredients and we focus on these fields in this
paper.  The Penrose transform for $H^3$s, which we will describe as the direct transform,  is most straightforward being easily obtained by an integral formula \eqref{direct-h3} and was the first to be 
studied in \cite{Hughston:1987, Hughston:1988nz,Hughston:1990, Berkovits:2004bw}.   The Penrose transform for $H^2$s is not direct, although we give an integral formula in \eqref{indirect-h2a}.  This is the case that naturally gives rise to   a holomorphic gerbe on twistor space \cite{Chatterjee} and so may well be the most geometrically natural for making contact with the $(0,2)$ gerbe theory.\footnote{We remind the reader that, loosely speaking, a gerbe is the extension of the concept of a line bundle with connection 1-form (or $(0,1)$-form deforming a $\dbar$-operator in the holomorphic case) replaced by a connection 2-form (or $(0,2)$-form in the holomorphic case).}   In \cite{Chern:2009nt} a
super-twistor space was introduced and some speculations were made concerning the Penrose transform of  the $(0,2)$-theory  (in $H^3$ form).

Much recent progress in the study of scattering amplitudes in four dimensions has exploited the simplicity arising from spinor helicity methods and these have recently been extended to six dimensions and higher \cite{Cheung:2009dc,Boels:2009bv}.  Spinor helicity is essentially a method for efficiently encoding the polarization information of
momentum eigenstates in terms of spinors.  It is a useful starting
point for exploring the twistor space formulation of scattering amplitudes and
thereby gives insight into the structure of the twistor formulation of the theory; it is straightforward to construct the twistor
space representation of an amplitude given the twistor representation
of momentum eigenstates and the momentum space amplitude in spinor
helicity form. This has been instrumental in recent progress, both for expressing known amplitudes in twistor space and for obtaining amplitudes on momentum space from their counterparts in twistor space, see for
example \cite{Witten:2004cp, Boels:2007qn,Mason:2008jy}.  

In this paper we obtain distributional Dolbeault representatives on twistor space for such spinor-helicity representations of momentum
eigenstates.  We do so in a way that naturally extends to higher dimensions and indeed gives some explanation of the structure
of the Penrose transform in higher dimensions; the fact that in dimension $2m$, the relevant cohomology is obtained in $H^{m-1}$ and
$H^{m(m-1)/2}$, and the fact that the helicities obtained of these different cohomology degrees are the same for $m$ odd, and opposite for $m$ even. In brief, the spinor-helicity data for a momentum eigenstate  consists of  a null momentum $P$ and polarization data $\xi$ in some irreducible representation of the `little group', i.e., the $\SO(2m-2)$ inside the stabilizer of $P$.   For our massless fields, the polarization data will be irreducible symmetric spinor representations of this little group and these can be obtained both as holomorphic sections of line bundles, or dually as $H^{(m-1)(m-2)/2}$s on the chiral projective pure spinor spaces for the little group.  These can then be combined with distributional Dolbeault forms supported on the little group spin space inside twistor space to generate the spinor-helicity Dolbeault representatives.  In four dimensions, with $m=2$, the projective pure spin space for the little group is trivial (just a point) and so this feature does not play a role, but in six it is the Riemann sphere and so this is the first nontrivial case where this comes in.  It is the dimension of this little group pure spin space that gives the difference between the cohomology  degrees in the Penrose transform.

These twistor representatives
associated to spinor helicity forms of the momentum eigenstates are a
useful tool that allow us to understand key features of the Penrose transform.  For example, they give an elementary proof of the Penrose transform isomorphism without the spectral sequences of \cite{Baston:1989vh}.
One feature that is peculiar to six dimensions is that the same massless fields are represented  both as $H^2$s and $H^3$s.  We will see that the best way to understand the correspondence  between these two representations is to consider the problem of extending the cohomology classes off $Q$ into $\CP^7$.  The $H^2$s turn out to have a unique extension to the order given by the helicity, but any further extension is obstructed with obstruction given by the corresponding $H^3$.  This understanding of the extrinsic behaviour of the cohomology class allows us to write an integral formula in this indirect case.
The proof  here is expedited by the spinor-helicity representatives introduced earlier.  Although such features are not encountered in four-dimensional twistor space $\CP^3$, they are reminiscent of the extensions off ambitwistor space that arise in the description of four-dimensional physics \cite{Witten:1978xx,Isenberg, Pool, Eastwood:1985,Eastwood:1986,Baston:1987av,Eastwood:1987} although in that case it is  $H^1$s that are extended with obstructions in $H^2$s as opposed to $H^2$s being extended with obstructions in $H^3$.
 
Another tool that has proved important in four dimension is the half-Fourier transform introduced by  Witten \cite{Witten:2003nn} and developed and exploited in \cite{Mason:2009sa}.   The Penrose transform is replaced here by an integral transform first studied by Sparling \cite{Sparling:2006zy,Sparling:2007} for the wave equation and referred to as the $\Xi$-transform.  In four dimensions this is known as the X-ray transform and takes functions on real twistor space, $\RP^3$ to functions on split signature space-time by integrating over lines.  This can then be concatenated with the Fourier transform which then gives functions on the momentum light-cone.  This is the half-Fourier transforms which involves Fourier transforming in two of the four non-projective variables.  In six dimensions twistor space is now six-dimensional whereas the momentum light-cone is only five-dimensional and so there must be some loss of information.  The spinor helicity momentum eigenstates have a natural representation in this split signature also, and lead to a natural generalization of the half-Fourier transform from functions on twistor space to the momentum light-cone augmented by the pure spinor space for the little group at each point.  This is also a six-manifold and hence can be, and is an isomorphism.   

 We make a preliminary investigation of amplitudes.  
However, for the self-dual gerbe theory, Huang and Lipstein
\cite{Huang:2010rn,arXiv:1110.2791} have observed, using the spinor-helicity
formalism, that there are no good candidates for amplitudes that are
local in space-time.  This is in keeping with the idea that the self-dual gerbe theory only exists as a strongly coupled theory in six dimensions with no parameters that  can be taken as a coupling constant to expand in to obtain perturbative amplitudes.  
Because we have restricted our attention to conformally invariant chiral fields in this paper, the only nontrivial interaction we are able to consider is 
the  basic conformally invariant $\phi^3$ vertex, but we see that it has a natural formulation on twistor space. 

 In this article, we find new features important to higher-dimensional
 twistor theory that are absent in the four-dimensional case. This may
 account for some of the remarkable features of six-dimensional
 quantum theories.  One of the motivations for studying
 higher-dimensional twistor space was the realisation that twistor
 theory is the natural framework to study extended supersymmetry
 \cite{Ferber:1977qx}. It is tempting to suggest that a study of
 higher-dimensional supersymmetric physics from a twistorial
 perspective may illuminate many of the mysterious properties that
 higher-dimensional supersymmetric theories exhibit
 \cite{Strominger:1995ac,Hull:2000zn,Schwarz:2000zg,Seiberg:1996vs,Witten:1995zh,Witten:1996hc,Ganor:1996nf,arXiv:1108.4060}.
 We hope to return to this in a subsequent paper.

The format of this paper is as follows. The next section gives an overview of the geometry of six-dimensional twistor space and the Penrose transforms that relate cohomology classes in twistor space to the solutions of zero-rest-mass field equations. Section three presents a higher-dimensional formalism for relating spinor-helicity methods and twistor theory that generalises the four-dimensional approach outlined in \cite{Witten:2004cp} to six dimensions. Although the focus is on the six-dimensional case, in section four we show how our formalism works in arbitrary dimension. Section five considers two integral transforms that exist in split-signature. The first is a six-dimensional analogue of the half-Fourier transform introduced by Witten in \cite{Witten:2003nn} which relates objects on null momentum space to objects in twistor space. The second is the $\Xi$-transform introduced by Sparling in \cite{Sparling:2006zy,Sparling:2007}. We find an analogue of the half-Fourier transform in six dimensions and extend the $\Xi$-transform to include fields of arbitrary spin.  We give a brief discussion of interacting theories in six-dimensions  focussing on the conformal $\Phi^3$ scalar theory. We show how the three-point amplitude for this theory may be constructed from six-dimensional twistor space and close  with some discussion  of the formulation of conformal field theories in twistor space.   In an appendix, we give the indirect Penrose tranform for a gerbe.

Whilst this preparing this manuscript we learnt of work \cite{SW} which overlaps with some sections of this article.

\section{The Six-Dimensional Twistor Correspondence}\label{sec-6dimTwistor}

In this section we review the structure of six-dimensional twistor space and the Penrose transform. For reviews of the four-dimensional case see \cite{Huggett:1986fs,Ward:1990vs,Adamo:2011cb}. In order to deal with a variety of different signatures, we will start by working on complexified Minkowski space, $\M^I=\C^6$. We can extend $\M^I$ to the compactified, complexified Minkowski space $\M$ by adding a lightcone at infinity to give a six-quadric $\M$ in $\CP^7$. 
For the most part, we shall be working some region in $\M$ or one of its real slices and its associated twistor space.  Introducing spinor indices from the start, we can coordinatize $\M^I$ with coordinates $x^{AB}=x^{[AB]}, A,B=1,\ldots,4$ equipped with metric 
$$
\rd s^2 =\frac12 \varepsilon_{ABCD}\rd x^{AB}\rd x^{CD}\, , \qquad \varepsilon_{ABCD}=\varepsilon_{[ABCD]}\, , \qquad \varepsilon_{0123}=1\, .
$$
The two four-dimensional chiral spinor representations, dual to one another, are denoted $\Ss^A$ and $\Ss_A$ with the given index structure and have structure group $\SL(4,\C)$ in the complex.  

The \emph{real} slice $\M^I_{p,q}$ of $\M^I$ are those $\R^6\subset \C^6$ on which the metric has signature $(p,q)$.\
On Euclidean, Lorentzian or split signature real slices, the spin group reduces to the  real subgroups $\SU(4)$, $\SL(2,\bbH) $ or $\SL(4,\R)$ respectively.   These can be characterized as follows
\begin{itemize}
\item $(p,q)=(6,0)\leadsto \SU(4)$, the subgroup of $\SL(4,\C)$ that commutes with the conjugation $\pi_A\rightarrow \bar\pi^A$ with $\pi_A\bar\pi^A$ positive definite.
\item $(p,q)=(5,1)\leadsto \SL(2,\bbH)$ is the subgroup commuting with a quaternionic conjugation $\hat{}:\pi_A \mapsto \widehat{\pi_A} =: \hat\pi_A $ on $\Ss^A$ (and on $\Ss_A$) where
$$
\hat\pi_A=(-\bar\pi_1,\bar\pi_0, -\bar\pi_3,\bar\pi_2)\, .$$
We have $\hat{\hat{\pi}}_A=-\pi_A$ so there are no real spinors.
\item $(p,q)=(4,2)\leadsto \SU(2,2)$,
the subgroup of $\SL(4,\C)$ that commutes with the conjugation $\pi_A\rightarrow \bar\pi^A$, but this time, $\Ss^A$ divides into three parts according to the definiteness of $\pi_A\bar\pi^A$.
\item $(p,q)=(3,3)\leadsto \SL(4,\R)$, the group commuting with a conjugation $\pi_A\rightarrow\bar\pi_A$ and we  can take spinors to be real $\pi_A=\bar\pi_A$.
\end{itemize}

There is no natural way to raise or lower individual spin indices; however, skew-symmetric \emph{pairs} of indices can be raised and lowered by means of $\frac{1}{2} \varepsilon_{ABCD}$, i.e. $v_{AB} = \frac{1}{2} \varepsilon_{ABCD} v^{CD}$. We have the useful identities $v_{A C} v^{B C} = \frac{1}{4} \delta \ind*{_A ^B} v^2$ for any vector $v^{AB}$.  

A null vector $v^{AB}$ satisfies $v^{\lb{A} B} v^{C \rb{D}} = 0$, and so has vanishing determinant as a $4\times 4$ matrix.  Since skew matrices have even rank,  if it is to be non-trivial, this rank must be two so it can  be decomposed as $v_{AB} = \lambda^0_{\lb{A}} \lambda^1_{\rb{B}}$ for some spinors $\lambda^a_A=(\lambda^0_A, \lambda^1_A)$. This decomposition is not unique, but is subject to the $\SL(2,\C)$ freedom $\lambda^a_A \mapsto  \Lambda^a{}_b\lambda_A^b$ for some $\Lambda^a{}_b\in\SL(2,\C)$.
For a real null vector in Lorentz signature we can take $\lambda^1_A=\hat\lambda^0_A$ and in split signature the $\lambda^a_A$ can be taken to be real.

\subsection{Twistor space}

Twistor space $Q$ is the six-dimensional projective quadric in $\CP^7$ \cite{Penrose:1986ca}.
We  introduce the homogeneous coordinates $Z^\alpha = ( \omega^A , \pi_A ) \in \Ss^A \oplus \Ss_A$ on $\CP^7$, and define the quadric as
\begin{align}\label{Q-defn}
Q:= \{ [Z^\alpha] \in \C\P^7 | Z^2= 0 \} \quad \mbox{ where }\quad Z^2= 2 \omega^A \pi_A \, .
\end{align}
The relationship with space-time $\M^I$ follows from the
incidence relations
\begin{align}\label{eq-incidence_relation+}
 \omega^A & = x \ind{^{A B}} \pi_{B} \, ,
\end{align}
where we assume $\pi_B \neq 0$ for now. Since $x^{AB}$ is
skew-symmetric, we must have the condition $\omega^A \pi_A = 0$, i.e., the quadric condition $Z^2=0$. 
Holding $x^{AB}$ fixed in equation \eqref{eq-incidence_relation+} and
varying $[\omega^A, \pi_A]$, this defines a three-dimensional linear subspace $\C\P^3\subset Q$, which we shall denote $S_x$.  The converse statement, what a point in twistor space corresponds to in space-time,  is found by holding $[\omega^A,\pi_A]$ fixed  
and allowing $x^{AB}$ to vary. This determines a totally null self-dual three-plane $x^{AB}=x_0^{AB}+\varepsilon^{ABCD}\pi_A\alpha_B$ for some $\alpha_B$ defined modulo $\pi_B$. This is an \emph{$\alpha$-plane}.

We have so far ignored the case when $\pi_A = 0$, but we can easily extend the geometric interpretation of twistor space to compact complex spacetime $\M$ by simply identifying $[\omega^A,0]$ as the $\CP^3$ corresponding to the point added to $\M^I$ at infinity. Notationally, $Q$ and $Q^I$ will refer to the twistor spaces of $\M$ and $\M^I$ respectively.

\begin{center}
\begingroup
  \makeatletter
  \providecommand\color[2][]{%
    \errmessage{(Inkscape) Color is used for the text in Inkscape, but the package 'color.sty' is not loaded}
    \renewcommand\color[2][]{}%
  }
  \providecommand\transparent[1]{%
    \errmessage{(Inkscape) Transparency is used (non-zero) for the text in Inkscape, but the package 'transparent.sty' is not loaded}
    \renewcommand\transparent[1]{}%
  }
  \providecommand\rotatebox[2]{#2}
  \ifx\svgwidth\undefined
    \setlength{\unitlength}{10cm}
  \else
    \setlength{\unitlength}{\svgwidth}
  \fi
  \global\let\svgwidth\undefined
  \makeatother
  \begin{picture}(1,0.70707072)%
    \put(0,0){\includegraphics[width=\unitlength]{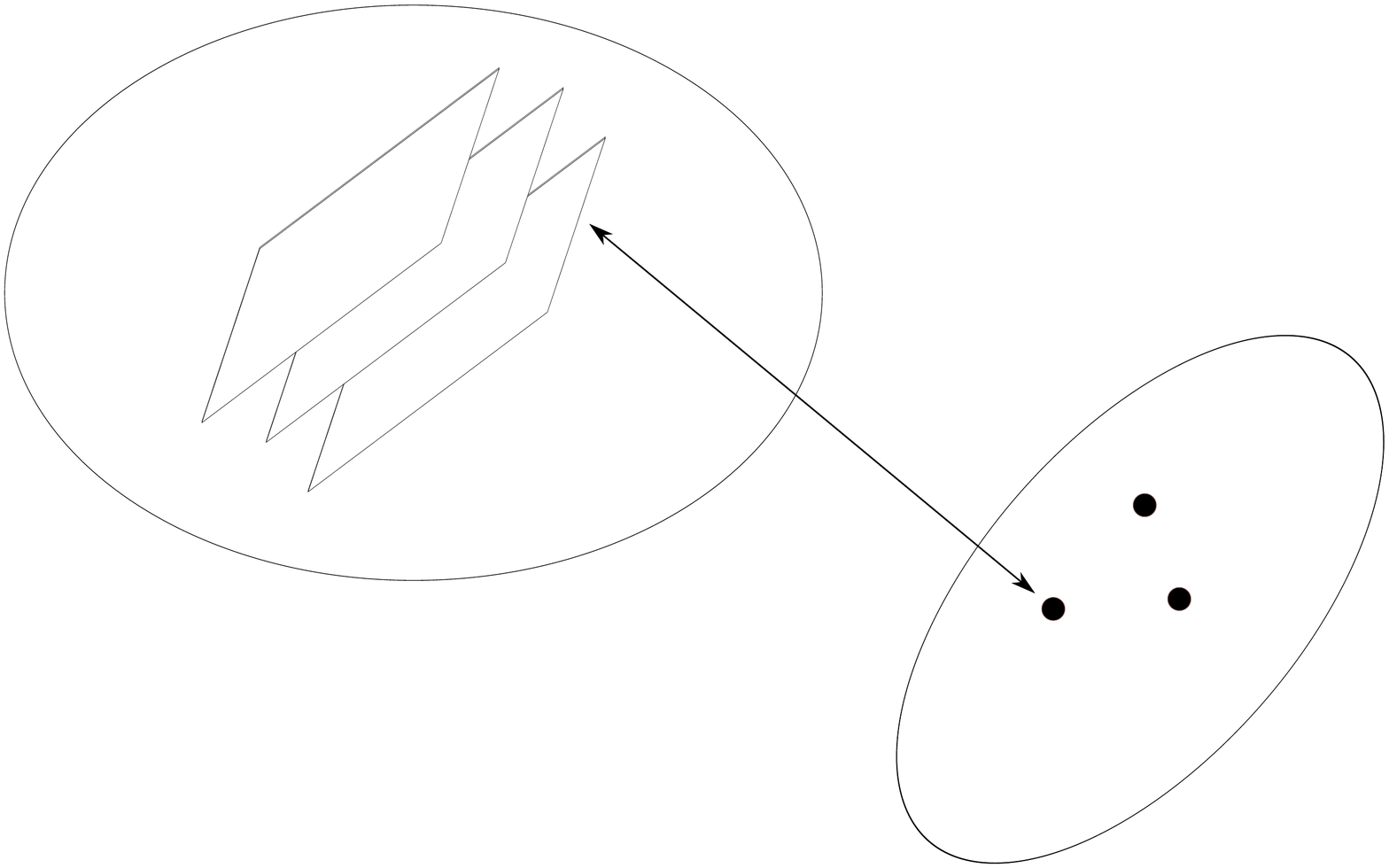}}%
    \put(0.68811121,0.23688535){\color[rgb]{0,0,0}\makebox(0,0)[lb]{\smash{$Z_3$}}}%
    \put(0.82421945,0.26744222){\color[rgb]{0,0,0}\makebox(0,0)[lb]{\smash{$Z_2$}}}%
    \put(0.75118122,0.35734271){\color[rgb]{0,0,0}\makebox(0,0)[lb]{\smash{$Z_1$}}}%
    \put(0.24429173,0.3131843){\color[rgb]{0,0,0}\makebox(0,0)[lb]{\smash{$\mathbb{M}$}}}%
    \put(0.36782331,0.56974998){\color[rgb]{0,0,0}\makebox(0,0)[lb]{\smash{$\alpha_2$}}}%
    \put(0.35198596,0.42879721){\color[rgb]{0,0,0}\makebox(0,0)[lb]{\smash{$\alpha_3$}}}%
    \put(0.1841096,0.5539126){\color[rgb]{0,0,0}\makebox(0,0)[lb]{\smash{$\alpha_1$}}}%
    \put(0.62932651,0.15583519){\color[rgb]{0,0,0}\makebox(0,0)[lb]{\smash{$Q\subset\C\P^7$}}}%
    \put(0.68811121,0.23688535){\color[rgb]{0,0,0}\makebox(0,0)[lb]{\smash{$Z_3$}}}%
    \put(0.82421945,0.26744222){\color[rgb]{0,0,0}\makebox(0,0)[lb]{\smash{$Z_2$}}}%
    \put(0.75118122,0.35734271){\color[rgb]{0,0,0}\makebox(0,0)[lb]{\smash{$Z_1$}}}%
    \put(0.24429173,0.3131843){\color[rgb]{0,0,0}\makebox(0,0)[lb]{\smash{$\mathbb{M}$}}}%
    \put(0.36782331,0.56974998){\color[rgb]{0,0,0}\makebox(0,0)[lb]{\smash{$\alpha_2$}}}%
    \put(0.35198596,0.42879721){\color[rgb]{0,0,0}\makebox(0,0)[lb]{\smash{$\alpha_3$}}}%
    \put(0.1841096,0.5539126){\color[rgb]{0,0,0}\makebox(0,0)[lb]{\smash{$\alpha_1$}}}%
\put(0.0,0.0){\color[rgb]{0,0,0}{\makebox(0,0)[lb]{\smash{\emph{Figure 1: Self-dual null planes are represented by points in twistor space.}}}}} 
  \end{picture}%
\endgroup
\end{center}

 In contrast to the four-dimensional case where a null line in space-time is a point in twistor space, null lines in six-dimensional space-time correspond to null lines in six-dimensional twistor space.
Generically two $\alpha$-planes, $\alpha_1$ and $\alpha_2$, do not intersect.   However, if $Z_1\cdot Z_2=0$, then they do and indeed they must do in a null line ${\cal L}=\alpha_1\cap\alpha_2$ \cite{Hughston:1986hb}.  In twistor space, this configuration is given by the line joining the two twistors, $Z_1$ and $Z_2$, connected by the null line ${\cal L}'$ corresponding to the null line ${\cal L}$ in space-time (if $Z_1\cdot Z_2\neq 0$, then the line joining $Z_1$ to $Z_2$  in $\CP^7$ will not lie in $Q$).

\begin{center}
\begingroup
  \makeatletter
  \providecommand\color[2][]{%
    \errmessage{(Inkscape) Color is used for the text in Inkscape, but the package 'color.sty' is not loaded}
    \renewcommand\color[2][]{}%
  }
  \providecommand\transparent[1]{%
    \errmessage{(Inkscape) Transparency is used (non-zero) for the text in Inkscape, but the package 'transparent.sty' is not loaded}
    \renewcommand\transparent[1]{}%
  }
  \providecommand\rotatebox[2]{#2}
  \ifx\svgwidth\undefined
    \setlength{\unitlength}{10cm}
  \else
    \setlength{\unitlength}{\svgwidth}
  \fi
  \global\let\svgwidth\undefined
  \makeatother
  \begin{picture}(1,0.70707072)%
    \put(0,0){\includegraphics[width=\unitlength]{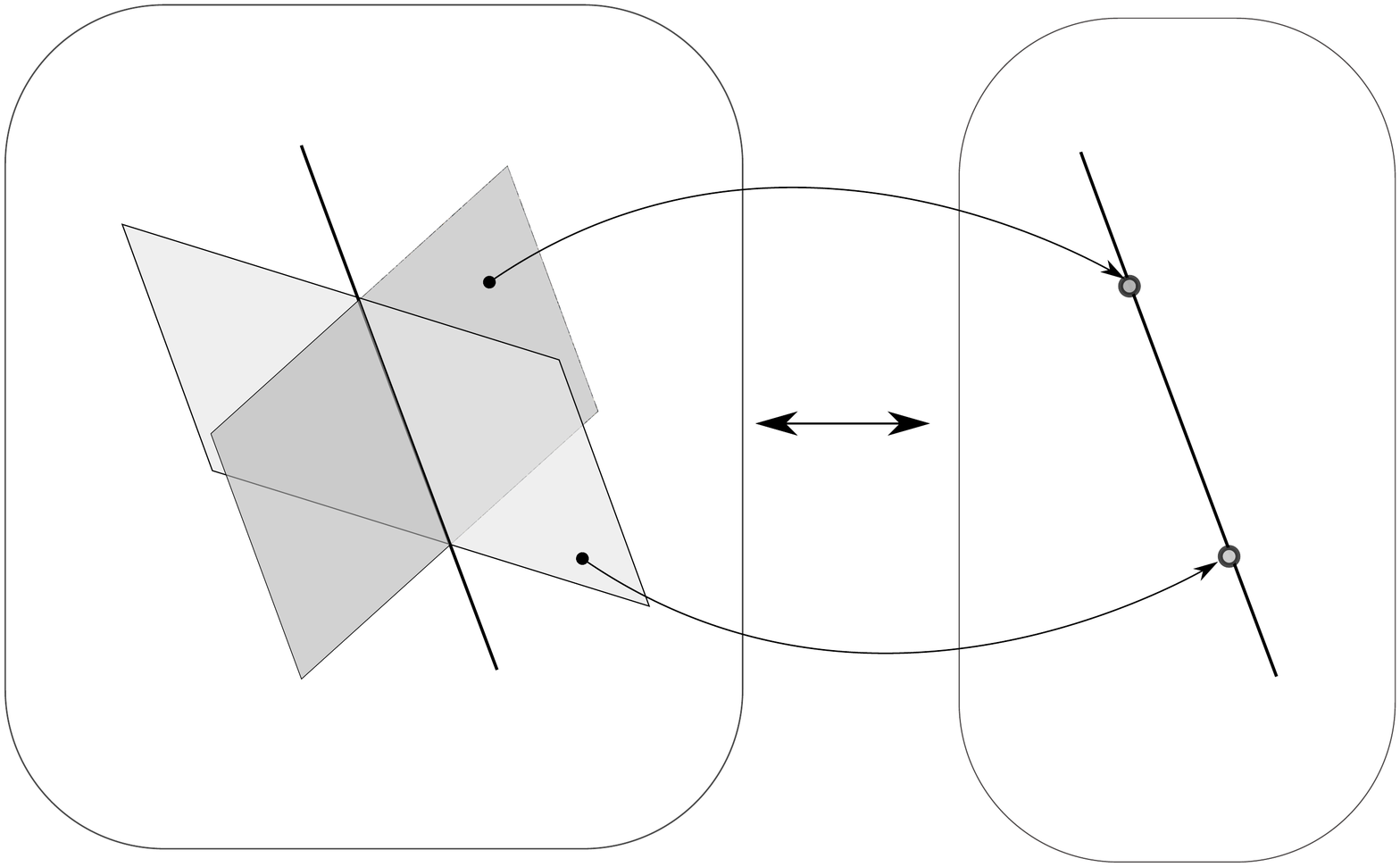}}%
    \put(0.83591706,0.45433503){\color[rgb]{0,0,0}\makebox(0,0)[lb]{\smash{$Z_1$}}}%
    \put(0.8938225,0.27171016){\color[rgb]{0,0,0}\makebox(0,0)[lb]{\smash{$Z_2$}}}%
    \put(0.7587098,0.56717641){\color[rgb]{0,0,0}\makebox(0,0)[lb]{\smash{${\cal L}'$}}}%
    \put(0.19895717,0.5642069){\color[rgb]{0,0,0}\makebox(0,0)[lb]{\smash{$\quad{\cal L}$}}}%
    \put(0.32909419,0.47957586){\color[rgb]{0,0,0}\makebox(0,0)[lb]{\smash{$\quad\alpha_1$}}}%
    \put(0.06384446,0.41127713){\color[rgb]{0,0,0}\makebox(0,0)[lb]{\smash{$\qquad\qquad\alpha_2$}}}%
\put(0.0,0.0){\color[rgb]{0,0,0}{\makebox(0,0)[lb]{\smash{\emph{Figure 2: Two $\alpha$-planes always intersect in a null line in $\mathbb{M}$.}}}}} 
  \end{picture}%
\endgroup
\end{center}

Twistor space is a quadric in the projectivation of the positive
chiral spinor representation $\Ss^\alpha$ of $\SO(8,\C)$.  Primed
twistor space $Q'$ is the quadric in the projectivation of the negative
chiral spinor representation $\Ss^{\alpha'}$ of $\SO(8,\C)$ which we
can coordinatize as $W^{\alpha'}=(\mu_A,\lambda^A)$ with inner product
$W\cdot W=2 \lambda^A\mu_A$. 
Primed twistor space $Q'$ is the space of all anti-self-dual null
planes, \emph{$\beta$-planes}, in $\M$ given by $\mu_A=x_{AB}\lambda^B$. Analogously to $Q^I$, the primed twistor space of $\M^I$ will be denoted ${Q'}^I$.

This primed twistor space is the six-dimensional analogue of the
\emph{dual twistor space} from four dimensions. The latter
is also a primed twistor space, but happens to be isomorphic to
the dual of twistor space,  a feature of dimensions $0
\mod 4$. However, in dimensions $2 \mod 4$, twistor space and
primed twistor space have canonically defined inner products and
so are dual to themselves. Thus important aspects of the Penrose transform
are different in these cases as we shall see.

As a final remark, we note that one of the features of six-dimensional twistor theory is
the \emph{triality} between the three six-dimensional complex quadrics
$\M$, $Q$ and $Q'$. This can be described by means of the generators
of the Clifford algebra for $\SO(8,\C)$, but here, it suffices to say
that the geometric correspondences between $\M$ and $Q$, and $\M$ and
$Q'$ are also mirrored by one between $Q$ and $Q'$, e.g. an
$\alpha$-plane in $Q'$ corresponds to a point in $Q$, and so on.
Triality holds in split signature, but is broken by other choices of
signature (it holds also for $\SO(8,\R)$ but then the real quadrics are empty).

\subsection{Massless fields}
 Our main interest in this article will be conformally invariant chiral theories. Unlike in four dimensions, Yang-Mills theories cannot be chiral in six dimensions nor are they conformally invariant.  We will instead focus on symmetric spinor fields $\phi_{A_1 \ldots A_k}=\phi_{(A_1 \ldots A_k)}$ and $\psi^{A_1 \ldots A_k}=\psi^{(A_1 \ldots A_k)}$  describing chiral fields of helicity $\frac{k}{2}$ and $-\frac{k}{2}$ respectively.   In linearised, or non-interacting theories, such fields with $k>0$ satisfy the zero-rest-mass equations of motion
\begin{align*}
 \nabla \ind{^{B A_1}} \phi_{A_1 A_2 \ldots A_k} & = 0 \, , & \nabla \ind{_{B A_1}} \psi^{A_1 A_2 \ldots A_k} & = 0 \, , \quad \mbox{where }\quad \nabla_{AB}=\frac{\p}{\p x^{AB}}\, .
\end{align*}
 For scalar fields ($k=0$) we have
\begin{align*}
 \Box \phi & = 0 \, ,
\end{align*}
For any $k \in \Z$, the space of massless fields of helicity $k/2$ on a region $U$ in space-time will be denoted $\Gamma_{k}(U)$ although we will often drop the dependence on $U$. 

\subsection{The Penrose transform}
In six dimensions, the Penrose transform \cite{Eastwood:1981jy,Baston:1989vh} relates the space of massless fields of non-negative helicity on a region $U\subset\M$ to the cohomology classes with values in certain holomorphic line bundles $\mcO(m)$ over the corresponding region $Q^U$ swept out by the $S_x$ with $x\in U$ in twistor space. For most of the paper, we shall in fact take $U$ to be $\M^I$, complexified space-time, so that the corresponding twistor space $Q^U$ will simply be $Q^I$, the six-dimensional quadric with a $\CP^3$ removed. The line bundles $\mcO(m)$  are the restriction of the corresponding line bundles from $\CP^7$ whose sections can be identified with holomorphic functions on the nonprojective space $\C^8$ that are homogeneous of degree $m$. We denote these cohomology classes by $H^{\bullet}(Q^U;{\cal O}(m))$.

We note that one can obtain massless fields of negative helicity from
cohomology of holomorphic line bundles over some subset of
\emph{primed} twistor space. This is in contradistinction to
four-dimensional twistor theory, where fields of any integral helicity
can be obtained from twistor space. This dichotomy corresponds to that
of whether twistor space is canonically isomorphic to itself or to its
primed counterpart.

There are two ways one can obtain $\Gamma_k$ for $k \geq 0$ from cohomology classes in $Q$.  The first, the \emph{direct
  Penrose transform} is
$$
{\cal P}:H^3(Q^U,{\cal O}(-k-4))\stackrel{\simeq}\longrightarrow \Gamma_k(U) \, .
$$
This is direct in the sense that it follows explicitly by means of an integral formula: 
\begin{align} \label{direct-h3}
F_{A_1...A_k}(x)=\int_{S_x} \pi_{A_1}...\pi_{A_k} f(x\cdot\pi,\pi)\D^3 \pi \, ,
\end{align}
where $f = f(\omega,\pi)$ is a $(0,3)$-form homogeneous of degree $-k-4$  representing the cohomology class\footnote{We focus on Dolbeault cohomology here but it is sometimes more elegant to express the cohomology class in terms \v Cech cohomology, by use of the \v Cech-Dolbeault isomorphism
$$
\check{H}^{\bullet}(Q;{\cal O}(m))\cong H^{\bullet}_{\bar{\partial}}(Q;{\cal O}(m))
$$
so that massless space-time fields can be described by holomorphic functions on regions in twistor space.  In this case \eqref{direct-h3} is interpreted as a contour integral surrounding poles of $f$.} on twistor space, and $\D^3\pi=\varepsilon^{ABCD}\pi_A\rd \pi_B\rd \pi_C\rd\pi_D$ is the projective volume form on $S_x=\C\P^3$. 

The second transform will be referred to as the \emph{indirect Penrose transform}
$$
\widetilde{\cal P}:H^2(Q^U,{\cal O}(k-2))\stackrel{\simeq}\longrightarrow \Gamma_k(U) \, .
$$
In this case, the cohomology classes in $H^2(Q^U,{\cal O}(k-2))$ have trivial restriction on $\CP_3$ embedded in $Q^U$. For this reason, one cannot obtain the value of a massless field directly. The transform most naturally yields a potential modulo-gauge description\footnote{We elaborate on this construction in section six.} of the field, but this is nevertheless equivalent locally to the given field.

\section{Spinor-helicity representatives} 
In this section we turn to the problem of how to relate the six-dimensional spinor helicity methods proposed in \cite{Cheung:2009dc} to the six-dimensional twistor theory discussed in the preceding section. We first review the spinor-helicity formalism.

\subsection{The spinor-helicity formalism}
For the fields $\Gamma_{k}$, a momentum eigenstate is a field
$\Phi_{A_1\ldots A_k}=\Phi^0_{A_1\ldots A_k}\e^{\ii P\cdot x}$ with
$\Phi^0_{A_1\ldots A_k}$ constant.  The massless field equation then
implies that $P^{BA_1}\Phi^0_{A_1\ldots A_k}=0$.  We can solve these
algebraic equations as follows.  Firstly, if $P^{AB}$ were invertible
then $\Phi^0_{A_1\ldots A_k}$ would have to vanish.  Thus its rank must be less than four for a
nontrivial solution.  However, a skew matrix necessarily has even
rank so that if $P$ is to be nonzero, it must have rank two.  This
certainly implies $P^2=0$ and furthermore means that we can write
$$
P_{AB}= \varepsilon _{ab}\lambda_A^a\lambda_B^b \, ,\quad a,b=0,1,
$$
for some $\lambda_A^a$.  We refer to $a,b$ as `little group' indices.  The little group is $\SU(2)\times
\SU(2)$, the maximally compact part of 
the stabilizer of $P$ in $\Spin(1,5)$.  The representations of this 
maximal compact sub-group constitute the possible polarization data for massless
fields.  The $a,b$ are indices for the first of these $\SU(2)$ factors
and we will introduce $a', b'$ indices below for the second. Little group indices will be raised and lowered by means of the invariant skew forms $\varepsilon^{ab}$ and $\varepsilon_{ab}$ respectively, satisfying $\varepsilon_{ac} \varepsilon^{bc}= \delta_a^b$.
The reduction from $\SL(2,\C)$ to $\SU(2)$ is effected by choosing $\lambda^1_A=\hat \lambda^0_A$.

It is easily seen from the definition of $\lambda^a_A$ that $P^{AB}\lambda_B^a=0$  and in fact
that $\lambda_A^a$, $a=0,1$  span the kernel of $P^{AB}$.  Thus
$\Phi^0_{A_1\ldots A_k}=\lambda_{A_1}^{a_1}\ldots \lambda_{A_k}^{a_k}
\xi_{a_1\ldots a_k} $ for some $\xi_{a_1\dots a_k}=\xi_{(a_1\ldots
  a_k)}$.  
The irreducible polarization information is therefore contained in the polarization spinor $
\xi_{a_1\ldots a_k}$  which is a spin $k/2$ representation of the  little
group (which is 
the spin group two dimensions down, here a four-dimensional spin group).
We therefore have the spinor-helicity representation of a momentum
eigenstate as   
$$
\Phi_{A_1\ldots A_k}(x)= 
\lambda_{A_1}^{a_1}\ldots \lambda_{A_k}^{a_k}
\xi_{a_1\ldots a_k} 
\e^{\ii P\cdot x} 
$$
with data $(\lambda_A^a,\xi_{a_1\ldots a_k})$.
Massless fields of the opposite chirality and of mixed
type can be treated similarly\footnote{For the massless fields of the
  opposite chirality we introduce $\widetilde\lambda^{Aa'}$ so that
$$
\Phi^{A_1\ldots A_k}(x)= \widetilde\lambda^{A_1}_{a'_1}\ldots \widetilde\lambda^{A_k}_{a'_k}
\xi^{a'_1\ldots a'_k} \e^{\ii P\cdot x} \, \quad   \mbox{ where }
\quad P^{AB}= \varepsilon _{a'b'}\widetilde\lambda^{Aa'}\widetilde\lambda^{Bb'} \, ,
\quad a',b'=0,1.
$$

Linearised gravity and Maxwell theory are of mixed type. Maxwell
theory is described by a photon with (traceless) field strength
$F^A_B$ satisfying $\nabla_{AB}F^B_C=0=\nabla^{AB}F_B^C$ with momentum
eigenstates
$$
F^A_B(x)=\widetilde{\lambda}^A_{a'}\lambda_B^b\,\zeta^{a'}{}_b\,\e^{\ii P\cdot x}\;,
$$
for some little-group spinor $\zeta^{a'}{}_b$.
Similarly, linearised gravity is described by a linearised Weyl tensor
$\Psi_{AB}^{CD}$ symmetric in each pair, subject to
$\nabla_{AB}\Psi^{BC}_{DE}=0=\nabla^{AB}\Psi_{BC}^{DE}$ with
momentum eigenstates 
$$
\Psi^{AB}_{CD}(x)=\widetilde{\lambda}^A_{a'}\widetilde{\lambda}^B_{b'}\lambda_C^c\lambda_D^d\,\zeta^{a' b'}{}_{cd}\,\e^{\ii P\cdot x}\;,
$$
for some little group spinor $\zeta^{a'b'}_{cd} =
\zeta^{(a'b')}_{(cd)}$.  The corresponding polarization data
$\zeta^{a'}{}_b$ and $\zeta^{a'b'}_{cd}$ has
little-group indices of both types.} but we will not consider them
here although we will return to these in \cite{MRT}; they
require the use of cohomology with values in a vector bundle rather
than a line bundle, and in the mixed case require also the breaking of
conformal invariance

\subsection{Twistor representatives for momentum eigenstates}

We now find canonical twistor cohomology classes for such momentum eigenstates.  
Using Dolbeault cohomology, the twistor representatives will be $\dbar$-closed $(0,3)$-forms \eqref{h3-spin-hel-rep} of weight $-k-4$ 
for the direct transform \eqref{direct-h3} or a $(0,2)$-form \eqref{h2-spin-hel-rep} of
homogeneity $k-2$ for the indirect case. We first establish some machinery and notation.

We can encode the polarization information as a cohomology class on a Riemann
sphere $\CP^1_u$ acted on by the unprimed $\SU(2)$ of the little group. We will use homogeneous
coordinates $u_a$, $a=0,1$, on $\CP^1_u$ (hence the subscript).  
The polarization information can be
expressed as a holomorphic function homogeneous of degree $k$ on  $\CP^1_u$ 
$$
\xi(u)= \xi_{a_1\ldots a_k}u^{a_1}\ldots u^{a_k}\in H^0(\C\P^1;\cO(k))\, .
$$
Alternatively, by Serre duality, there is an $\alpha_\xi \in H^1(\CP^1, \cO(-2-k))$ such that  
$$
\xi_{a_1\ldots a_k}= \int_{\CP^1_u} u_{a_1}\ldots u_{a_k}
\alpha_\xi\wedge \D u\, , \quad \mbox { e.g. } \quad \alpha_\xi=
\xi_{a_1\ldots a_k}\hat u^{a_1}\ldots \hat u^{a_k}\,  \D \hat u\, ,
$$
where
$$
\hat u^a=\frac{1}{\left(|u_0|^2+|u_1|^2\right)}\left(\begin{array}{c}
\bar u_0\\
\bar u_1
\end{array}\right)\, ,\quad
\mbox{ and }\quad \D u =u_a\rd u^a\, , \quad \D\hat u=\hat u_a\rd \hat
u^a.
$$  
Here the $\hat{u}_a$ coordinates can be also invariantly defined as $\hat{u}_a = \frac{\bar{u}_a}{u_b \bar{u}^b}$, where $\bar{u}_a = (-\bar{u}_1,\bar{u}_0)$ are antiholomorphic coordinates on $\CP^1_u$, so that it is normalised, i.e. $u_a \hat{u}^a =1$, and is thus homogeneous of weight $-1$ in $u_a$.

The little group spinor $u$ defines two full spinors
$$
\lambda_A(u):=\lambda_A^au_a\, , \quad \widehat\lambda_A (u):= \lambda^a_A\hat u_a
$$
and we will have that $P_{AB}=\lambda_{[A}\widehat\lambda_{B]}$. We will
often suppress the indices and dependence on $u$ just writing $\lambda$ and $\widehat\lambda$ for these quantities in what follows and make the dependence explicit with $\lambda\cdot u$ where confusion might otherwise arise.

The twistor representative associated to the momentum space
eigenstate of null momentum $P$, will turn out to be supported on the line in
the `$\pi_A$' spin space spanned by the two $\lambda_A^a$, $a=1,2$.
This can be done with holomorphic delta functions on $\CP^3$.  To define these, we first introduce, for $z=x+iy$ 
$$
\bar \delta (z) = \delta( x)\delta(y) \rd \bar z \, .
$$
These can be multiplied together to give delta functions on $\C^4$: $\bar\delta^4(\mu_A)=\prod_{A=1}^4\bar\delta (\mu_A)$.  To give a delta function on the projective space $\CP^3$, for two points with  homogeneous coordinates $\mu_A$,
$\nu_A$ we define 
$$
\bar \delta^3_k(\mu,\nu)=\int_\C \frac{\rd s}{s^{k+1}}\wedge \bar \delta^4
(\mu + s \nu) 
$$
which has weight $-k-4$ in $\mu$ and $k$ in $\nu$.

Given spinor-helicity data $(\lambda_A^a,\xi_{a_1\ldots a_k})$, we can
now write the two formulae for  the twistor representatives. We
define the corresponding representative $\phi\in H^3(Q;{\cal
  O}(-k-4))$ to be 
\begin{equation}\label{h3-spin-hel-rep0}
\phi(\omega,\pi)=\e^{\ii P\cdot x} \,\int_{\CP^1_u} \alpha_\xi \wedge\bar
\delta^3_k(\pi,\lambda)\wedge  
\D u \, , 
\end{equation}
and the representative $\psi\in H^2(Q;{\cal O}(k-2))$ to be
\begin{equation}\label{h2-spin-hel-rep0}
\psi(\omega,\pi)=\e^{\ii P\cdot x} \, \int_{\CP^1_u} \, \xi( u) \, 
\bar \delta^3_{-2-k}(\pi,\lambda\cdot u)   \wedge \D u \, .
\end{equation}
Throughout, we take $k\geq 0$. 

In these formulae, $P\cdot x$ is not manifestly a twistor function; however, on the support of the delta function it is in the
sense that it satisfies $\pi_A\p^{AB} 
(P\cdot x)=0$ as $P^{AB}\pi_B=0$.  We can make this more explicit by
observing that 
$P_{AB}=\lambda_{[A} \widehat\lambda_{B]}$ and that on the support of the delta-function $\bar\delta^4(s\pi -\lambda)$
we can take $s\pi_A=\lambda_A$.  Thus, using the incidence relations
$x\cdot P=-s\omega\cdot\widehat\lambda$.  We therefore obtain as our definitive formulae for our twistor representatives of momentum eigenstates
\begin{eqnarray}
\phi(\omega,\pi)=\int_{\C\times\CP^1_u} \e^{- \ii s \omega\cdot\widehat\lambda}
\, \, \alpha_\xi \wedge\bar 
\delta^4(s\pi-\lambda\cdot u)\wedge  
s^{k+3}\rd s \wedge \D u \, ,  \label{h3-spin-hel-rep}
\\
\psi(\omega,\pi)=\int_{\C\times \CP^1_u} \e^{- \ii s\omega\cdot\widehat\lambda}  \, \, \xi( u) \, 
\bar \delta^4(s\pi-\lambda\cdot u)   \wedge \frac{\rd s}{s^{k-1}}\wedge \D u \, . \label{h2-spin-hel-rep}
\end{eqnarray}    
In both cases it is easy to check that all the weights and form
degrees match appropriately.  The $\dbar$-closure on $Z\cdot Z=0$ can be seen by checking $\dbar$-closure of the integrands of \eqref{h2-spin-hel-rep} and \eqref{h3-spin-hel-rep} as $\dbar$-closure is preserved by integration. In the $H^3$ case $\dbar$-closure follows by virtue of its holomorphic dependence on $\omega$ and  being a form of maximal degree on the $\CP^3\times \CP^1$ parametrised by $(\pi,u)$.  In the $H^2$ case, $ \xi( u) \, 
\bar \delta^3_{-2-k}(\pi,\lambda\cdot u)   \wedge \D u $ is
$\dbar$-closed.  Taking $\dbar $ of the exponential factor we obtain,
on the support of the delta function,
\be{dbar-closure}
 \dbar \big(\omega\cdot \widehat\lambda\big)= 
\omega^A\lambda_A^a\dbar
\hat u_a=-\omega^A\lambda_A^au_a \D \hat u 
\ee
where we have used the fact that $\dbar \hat u_a=u_a \D\hat u$.  On the
support of the delta function $\omega^A\lambda_A^a u_a=\omega\cdot\pi$
and so (\ref{dbar-closure}) vanishes on $Q$.

\section{Twistor spinor helicity states in higher dimensions}
The spinor helicity formalism and the  corresponding twistor space
representatives that we have presented in six dimensions have
straightforward extensions to higher dimensions.     One new issue that we meet in higher dimensions is the fact that
purity conditions come in for spinors in dimensions greater than six and symmetric spinors are then no longer an irreducible representation of the Lorentz group so that further irreducibility conditions must be imposed.  The other is that the canonical structures and identifications amongst spinors changes from dimension to dimension.  We will work in the complex to avoid further changes in character between the various signatures. 

The purity condition on a spinor $\pi^{A'}$ can be expressed
in many ways, but the most convenient for our purposes will be as
follows.  We first establish notation.  Let $\mu=1,\ldots , 2m$ be the
space-time indices, and 
$A=1,\ldots ,2^{m-1}$ and $A'=1',\ldots, 2^{m-1\prime}$ be the primed
and the unprimed spinor indices and decompose 
the gamma matrices into their chiral  parts, so that 
\be{}
\Gamma_{\mu}=\begin{pmatrix}  0&\gamma \ind{_{\mu A}^{B'}} \\ \gamma \ind{_{\mu A'}^B}&
  0\end{pmatrix}\, , 
\ee
and the Clifford algebra relations become
\be{}
 \gamma \ind{_{(\mu|A'}^{A}} \gamma \ind{_{|\nu)A}^{B'}} = - \eta_{\mu\nu}\,  \delta_{A'}^{B'}\, ,
 \qquad  \gamma \ind{_{(\mu|A}^{A'}} \gamma \ind{_{|\nu)A'}^{B}} = - \eta_{\mu\nu}\, \delta_{A}^{B}
\, ,
\ee
where $\eta_{\mu\nu}$ is the metric on $\C^{2m}$.
The purity condition on a spinor $\pi_A$ is the condition 
\be{purity}
\pi^{A'} \pi^{B'} \gamma \ind{_{\mu A'}^A} \gamma\ind{^{\mu}_{B'}^B} = 0 \, .
\ee
This guarantees that
the vector fields of the form $V^{\mu}=\gamma \ind{^\mu _{A'}^{A}} \pi^{A'} \alpha_A $ for
arbitrary $\alpha_{A}$ span
a totally null plane which can be of dimension at most $m$ and will, for non-zero $\pi^{A'}$, be $m$-dimensional.
 
\subsection{Spinor-helicity in higher dimensions}

We shall only be interested in the elementary conformally invariant symmetric spinor fields that we have been discussing
in six dimensions; massless fields such as Maxwell and linearized gravity are never conformally invariant in dimensions greater than four with their standard second order field equations. Thus we will take our massless fields to be 
symmetric spinors $\phi^{A_1'\ldots A_k'}=\phi^{(A_1'\ldots A_k')}$ and $\psi_{A_1\ldots A_k}=\psi_{(A_1\ldots A_k)}$. This choice allows us to deal with both cases $m$ even and odd at once. But when $m$ is odd, there will be some redundacy in this notation since primed indices can be eliminated by means of the isomorphism between primed spinor space and dual spinor space. This is consistent with the fact that the Penrose transform, as we describe it in this paper, produces massless fields of both positive and negative helicities when $m$ is even, but only massless fields of positive helicity when $m$ is odd. Further distinctions between these cases will be pointed out in the course of this section and the next. In dimensions greater than six these spinors are also subject 
to a further irreducibility condition.  This can be expressed in the form
\be{irred}
\phi^{A_1'\ldots A_k'}\gamma \ind{_{\mu A_1'}^{A}} \gamma \ind{^\mu_{A_2'}^{B}}=0\, , \qquad \psi_{A_1\ldots A_k}\gamma \ind{_{\mu A'}^{A_1}} \gamma \ind{^\mu_{B'}^{A_2}}=0\, .
\ee
The zero-rest-mass equations on such fields are then 
\be{higher-ZRM}
\gamma\ind{_{\mu A_1'}^{A}} \nabla^{\mu}\phi^{A'_1\ldots A'_k}=0\, , \qquad \gamma\ind{_{\mu A'}^{A_1}} \nabla^{\mu} \psi_{A_1\ldots A_k}=0\, .
\ee

We can now obtain the spinor helicity formalism for such fields.
We shall assume that our momentum eigenstates take the form $\phi^{A_1'\ldots A_k'}=\e^{\ii P \cdot x}\phi\ind*{^{A'_1\ldots A'_k}_0}$
and $\psi_{A_1\ldots A_k}=\e^{\ii P \cdot x} \psi\ind*{_{A_1\ldots A_k}^0}$ with $\phi^{A'_1\ldots A'_k}_0$ and $\psi\ind*{_{A_1\ldots A_k}^0}$ constant, and so
\be{}
P^{\mu}\gamma\ind{_{\mu A_1'}^{A}} \phi^{A'_1\ldots A'_k}=0\, , \qquad P^{\mu}\gamma\ind{_{\mu A'}^{A_1}} \psi_{A_1\ldots A_k}=0\, .
\ee
The Clifford algebra relations imply that $P^{\mu}\gamma\ind{_{\mu A'}^{B}}$ is
invertible unless $P$ is null in which case it is standard\footnote{This follows by choosing
another null vector $Q$ with $P\cdot Q=\frac{1}{2}$ and observing that $P \cdot
\gamma \cdot Q\cdot\gamma + Q\cdot
\gamma \cdot P\cdot\gamma = 1$ and this algebra has a standard
representation with 
$$
P\cdot\gamma=\begin{pmatrix} 0&I\\ 0& 0\end{pmatrix} , \qquad
Q\cdot\gamma=\begin{pmatrix} 0& 0\\ I& 0\end{pmatrix} \, ,
$$ 
where $I$ is the identity matrix on $\C^{2^{m-2}}$.} that
$P^{\mu}\gamma\ind{_{\mu A'}^{B}}$ is nilpotent with rank $2^{m-2}$. We can, as before, introduce bases $\lambda_a^{A'}$ and $\lambda^a_A$, ($a=1\ldots,2^{m-2}$), respectively, of the 
kernel of $P\cdot  \gamma$ and deduce that we must have
\be{}
\phi_0^{A'_1\ldots A'_k}=\xi^{a_1\ldots a_k}\lambda_{a_1}^{A'_1}\ldots
\lambda_{a_k}^{A'_k}\, , \qquad \psi^0_{A_1\ldots A_k}=\eta_{a_1\ldots a_k} \lambda^{a_1}_{A_1}\ldots
\lambda^{a_k}_{A_k}\, ,
\ee
for some symmetric $\xi^{a_1\ldots a_k}$ and $\eta_{a_1\ldots a_k}$.
However, we must now also implement the irreducibility conditions
\eqref{irred} on
$\xi^{a_1\ldots a_k}$ and $\eta_{a_1\ldots a_k}$. To do this we note that $a$ is a spinor
index for the group $\SO(2m-2,\C)$, the spin group for the space-time of two dimensions lower, which is the semi-simple part of the
stabilizer of $P$ acting on $P^\perp/P$.   The irreducibility conditions
\be{little-irred}
\xi^{a_1 a_2 \ldots a_k}\lambda_{a_1}^{A_1'}
\lambda_{a_2}^{A_2'} \gamma \ind{_{\mu A_1'}^{A_1}} \gamma \ind{^\mu _{A_2'}^{A_2}}=0 \, , \qquad
\eta_{a_1 a_2 \ldots a_k}\lambda^{a_1}_{A_1}
\lambda^{a_2}_{A_2} \gamma \ind{_{\mu A_1'}^{A_1}} \gamma \ind{^\mu _{A_2'}^{A_2}}=0
\ee
are then simply the analogues of \eqref{irred} for symmetric spinors $\xi^{a_1\ldots a_k}$ and $\eta_{a_1\ldots a_k}$ for
$\SO(2m-2,\C)$.  This is first non-trivial for dimension $2m=10$ when it is simply a trace-free condition on $\xi_{a_1\ldots a_k}$.
Here again, we remark that when $m$ is odd, the little group spinor space is isomorphic to its dual, and the little group spinor indices can be raised and lowered. Thus, in this case, the identification of $\lambda_a^{A'}$ with $\lambda^a_A$ is consistent with the fact that only massless fields of positive helicity are treated here. On the other hand, when $m$ is even, the little group spinor space and its dual are not isomorphic to one another, but correspond to distinct chiral spinor spaces for $\SO(2m-2,\C)$

Thus we are left with momentum eigenstates for solutions to \eqref{irred} and \eqref{higher-ZRM} given by the formula
\be{higher-spin-hel}
\phi^{A'_1\ldots A'_k}=\xi^{a_1\ldots a_k}\lambda_{a_1}^{A'_1}\ldots
\lambda_{a_k}^{A'_k} \e^{\ii P\cdot x}\, , \qquad \psi_{A_1\ldots A_k}=\eta_{a_1\ldots a_k}\lambda^{a_1}_{A_1}\ldots
\lambda^{a_k}_{A_k} \e^{\ii P\cdot x}\, .
\ee
This gives the chiral spinor-helicity description for these fields. 

\subsection{Twistor representatives for spinor-helicity states}

Following the last chapter of \cite{Penrose:1986ca}, in arbitrary even dimension $2m$ twistor space will be defined to be the projective pure spinors for the the conformal group $\SO(2m+2,\C)$. Here we work in the complex and make no restriction on $m$, so that we have to work independently of the special structures that arise in different dimension modulo 8. The space of projective positive pure spinors for $\SO(2m,\C)$ will be denoted $\PPS_m$. The space $\PPS_m$ has dimension $m(m-1)/2$ and can be represented as the space of $\alpha$-planes, the  homogeneous space $\SO(2m)/\U(m)$ or the complex subvariety of the ($2^{m-1}-1$)-dimensional projective spin space $\PS_m$ cut out by the purity conditions 
\begin{align*}\tag{\ref{purity}}
\pi^{A'} \pi^{B'} \gamma \ind{_{\mu A'}^{A}} \gamma \ind{^\mu _{B'}^{B}} =0\, .
\end{align*}
This condition guarantees that $V^{\mu}= \gamma \ind{^\mu _{A'} ^A}\pi^{A'} \alpha_A$ is a null vector for any choice of $\alpha_A$ and, with the purity condition, this will be maximal so that $\pi^{A'}$ will determine a totally null self-dual $m$-plane through the origin.  We take $\PT:=\PPS_{m+1}$ as our definition of (projective) twistor space, the space of such totally null self-dual $m$-planes but not necessarily through the origin in $\C^{2m}$.  For example, in dimension six ($m=3$) the twistor space is $\PT=Q=\PPS_4$, the space of projective pure spinors in eight dimensions. These null self-dual planes, often called $\alpha$-planes in this context, can be represented by the incidence relation
\be{higher=inc}
\omega^{A} = x^{\mu} \gamma \ind{_{\mu A'}^A} \pi^{A'} \, ,
\ee
as is familiar in dimension four and as was seen for dimension six in (\ref{eq-incidence_relation+}) where primed indices are eliminated in favour of the unprimed ones.  However, we note that \eqref{higher=inc} will be inconsistent unless $(\omega^A,\pi^{A'})$ is a pure $\SO(2m+2)$ spinor.

 If we now wish to proceed analogously to our development of the six-dimensional theory presented in previous sections, we must describe the physical degrees of freedom in terms of cohomology representatives on 
spaces of projective pure spinors, i.e. on twistor spaces.   Let $\PPS_{m-1}$ be the
$(m-1)(m-2)/2$-dimensional space of
projective pure spinors for $\SO(2m-2,\C)$.  We will take $u^a$ to be
homogeneous coordinates on the projective spin space $\PS_{m-1}=\CP^{2^{m-2}-1}$ and impose the
purity conditions
$$
u^{a_1} u^{a_2}\lambda_{a_1}^{A'_1}
\lambda_{a_2}^{A'_2} \gamma \ind{_{\mu A_1'}^{A}} \gamma \ind{^\mu _{A_2'}^{B}} =0\, .
$$
It is a standard consequence of Bott-Borel-Weyl theory \cite{Baston:1989vh} that the representations 
defined by \eqref{little-irred} can be represented by
\be{BBW}
 \alpha_\xi\in H^{\mathrm{top}}(\PPS_{m-1},\cO(4-2m-k)) \qquad \mbox{ and }\qquad \eta(u)\in H^0(\PPS_{m-1}^*,\cO(k))
\ee
 respectively, where top$=(m-1)(m-2)/2$ is the dimension of $\PPS_{m-1}$; these are related by Serre duality as the canonical bundle on $\PPS_{m-1}$ is $\cO(4-2m)$ (see \cite{Baston:1989vh}).
The second of these is simply given by $\eta(u)=\eta_{a_1\ldots a_k}u^{a_1}\ldots u^{a_k}$ whereas the first will not in general have a canonical representative, and we will denote such a representative simply by 
$\alpha_\xi$ satisfying
\begin{align*}
 \xi^{a_1 a_2 \ldots a_k} & = \int_{\PPS_{m-1}} u^{a_1} u^{a_2} \ldots u^{a_k} \alpha_{\xi} \D u \, ,
\end{align*}
where $\D u$ is the projective holomorphic volume form of weight $2m-4$ on $\PPS_{m-1}$. When $m$ is odd, one has $\xi_{a_1 \ldots a_k} = \eta_{a_1 \ldots a_k}$ since the little group spinor space is isomorphic to its dual, and we thus have two ways of representing the polarisation spinor of a given massless field of positive helicity.

We must also introduce the projective holomorphic volume form $\D\pi$ of weight $2m-2$ on $\PPS_m$ and the weighted delta function $\bar\delta(\pi,\rho)\in \Omega^{0,\frac{m(m-1)}2}(2-2m-k)$ on $\PPS_m$.  The former exists simply via the identification of the canonical bundle of $\PPS_m$ as the restriction of $\cO(2-2m)$ as follows from the Bott Borel Weil theory as described in \cite{Baston:1989vh} (explicit formulae in terms of the ambient projective coordinates in $\PS_m$ are obtained in \cite{Berkovits:2004bw}).   The delta function is defined tautologically by
\be{higher-delta}
f(\pi)=\int f(\rho)\, \bar\delta(\pi,\rho) \,\D \rho
\ee
where $f$ a function on $\PPS_m$ of weight $k$, $\D \rho$ is the canonical holomorphic volume form of weight $2m-2$ and $\bar\delta(\pi,\rho)$ has appropriate weights in each of its arguments for the formula to make sense, i.e., of weight $k$ in $\pi$ and weight $2-2m-k$ in $\rho$.  

We can now define the twistor representatives for the spinor-helicity states \eqref{higher-spin-hel} as
\begin{eqnarray}\label{higher-twis-spin-hel}
\phi(\omega,\pi)&=&\int \e^{\ii P \cdot x} \alpha_\xi\wedge \bar\delta(\pi,\lambda\cdot u)\, \D u  \nonumber \\
\psi(\omega,\pi)&=&\int \e^{\ii P\cdot x} \eta(u) \, \bar\delta(\pi,\lambda\cdot u)\, \D u 
\end{eqnarray}
and it can be seen that these are respectively $\dbar$-closed $(0,\frac{1}{2}m(m-1))$-forms of weight $2-2m-k$ and $(0,m-1)$-forms of weight $k-2$.  This follows from the weights $k$ of $\eta(u)$, $4-2m-k$ of $\alpha_\xi$ and  $2m-4$ of the canonical holomorphic volume form $\D u$ on $\PPS_{m-1}$.  It can be checked that $\e^{\ii P \cdot x}$ is indeed a function on twistor space when restricted to the support of the delta function as before as it will be annihilated by $\pi^{A'}\gamma \ind{^{\mu}_{A'}^A} \p_\mu$ on the support of the delta function.

\section{Formal neighbourhoods and the twistor transform}

In four dimensions a field of a given helicity is represented by the $H^1(\P\T^I;{\cal O}(m))$ cohomology classes on twistor space and the $H^1({\P\T^*}^I;{\cal O}(m-4))$ classes on dual twistor space.  The direct map between these two representations is known as the twistor transform.  In six dimensions, the situation is rather different as the fields of positive helicity $k/2$ correspond to classes on the same twistor space either as an $H^2$ of homogeneity $k-2$ or an $H^3$ of homogeneity $-k-4$. There is no description of such fields simply in terms of homogenous functions on dual twistor space.  We will identify the direct map between these representatives, the twistor transform, ${\cal T}$,  on twistor space in this section via an obstruction to the problem of extending the $H^2$s off the quadric $Q\subset \C\P^7$. The twistor transform can be written schematically as
\begin{align*}
\xymatrix{& \Phi_{A_1\dots A_k}(x) \ar[dl]_{{\cal P}^{-1}} \ar[dr]^{\widetilde{\cal P}^{-1}} & \\
g(Z) & \ar[l] {\cal T}\ar[r] & f(Z) }
\end{align*}
where $\Phi_{A_1\dots A_k}(x)\in\Gamma_k$, $g(Z)\in H^3(Q^I;(-k-4))$ and $f(Z)\in H^2(Q^I;(k-2))$.

The obstruction problem can also be motivated by the task of writing integral formulae for the indirect transform.  In four dimensions, there are integral formulae for the space-time fields associated to classes in $H^1(\P\T^I;{\cal O}(k-2))$ involving derivatives of the twistor function, for example at $k=2$, the self-dual photon, we have 
$$
F_{a'b'}(x)=\int_{\C\P^1_x} \D\lambda\,\frac{\p^2}{\p \mu^{a'}\p \mu^{b'}}\,f(x\cdot\lambda,\lambda)
$$
where $f\in H^1(\P\T^I;{\cal O})$.   Such a formula was proposed in dimension six  \cite{Berkovits:2004bw} with $(\mu,\lambda)$ replaced by $(\omega,\pi)$.  However, there are a number of problems:   in six dimensions, the cohomology classes are, a priori, only defined on $Q$ rather than $\CP^7$, and straightforward differentiation with respect to
$\omega^A$ takes a derivative in directions off $Q$ into $\CP^7$ and so are not immediately meaningful.  We will however show in the
following that certain such derivatives can be canonically defined in the positive homogeneity case (although there are other problems with the formula in \cite{Berkovits:2004bw} as the homogeneity weights and cohomology degrees are not right either).  

We will therefore consider the task of constructing extensions of the cohomology classes off the quadric as an expansion in powers of $Z^2$. In the $H^3$ cases we will see that classes can be extended off the quadric to all orders in $Z^2$, but that there is no way to fix the ambiguity that arises at each order and so the derivatives do not have any invariant meaning. In the $H^2$ case, we will show
\begin{propn} Every  $f\in H^2(Q^I,\cO(k-2))$ has a canonical  extension to the $k$th order formal neighbourhood around $Q$, so that in particular, its  $k$th derivative transverse to $Q$ is canonically defined; however, any further extension is obstructed, the obstruction being  the corresponding $g\in H^3(Q^I,\cO(-4-k))$ to which it corresponds by the twistor transform. 
\end{propn}
 Thus in examining the obstruction theory we find the canonical map from $H^2(Q^I,\cO(k-2))$ to $H^3(Q^I,\cO(-4-k))$ that we know must exist via the Penrose transform to space-time, intrinsically in twistor space.   This also enables us to  write integral formulae for the $H^2$ case.
 
Such extensions will be examined explicitly below for our representatives above. We will first examine the problem  abstractly of extending cohomology classes to formal neighbourhoods of the quadric $Z\cdot Z=2 \omega^A\pi_A=0$.  The subsequent explicit calculations will then demonstrate that certain maps are indeed isomorphisms as claimed.

\begin{center}
\begingroup
  \makeatletter
  \providecommand\color[2][]{%
    \errmessage{(Inkscape) Color is used for the text in Inkscape, but the package 'color.sty' is not loaded}
    \renewcommand\color[2][]{}%
  }
  \providecommand\transparent[1]{%
    \errmessage{(Inkscape) Transparency is used (non-zero) for the text in Inkscape, but the package 'transparent.sty' is not loaded}
    \renewcommand\transparent[1]{}%
  }
  \providecommand\rotatebox[2]{#2}
  \ifx\svgwidth\undefined
    \setlength{\unitlength}{8cm}
  \else
    \setlength{\unitlength}{\svgwidth}
  \fi
  \global\let\svgwidth\undefined
  \makeatother
  \begin{picture}(1,0.70707072)%
    \put(0,0){\includegraphics[width=\unitlength]{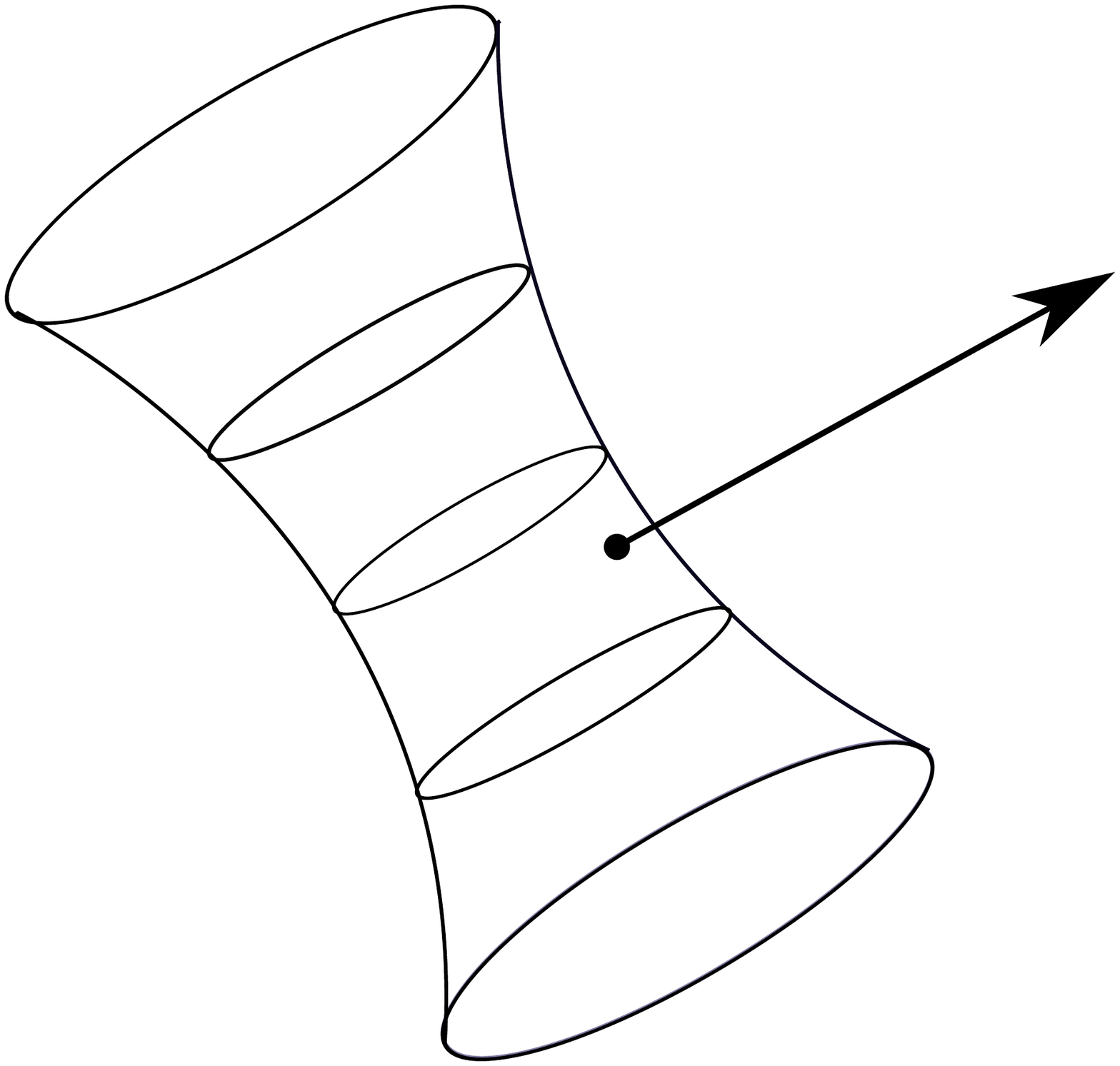}}%
    \put(0.67353026,0.52773915){\color[rgb]{0,0,0}\makebox(0,0)[lb]{\smash{$\bar{\p}f=Z^{2k+2}g$}}}%
    \put(0.83151573,0.28227145){\color[rgb]{0,0,0}\makebox(0,0)[lb]{\smash{$\C\P^7$}}}%
    \put(0.1800317,0.56346508){\color[rgb]{0,0,0}\makebox(0,0)[lb]{\smash{$Q$}}}%
    \put(0.30818227,0.31813144){\color[rgb]{0,0,0}\makebox(0,0)[lb]{\smash{$\bar{\p}f=0$}}}%
 \put(0.0,0.0){\color[rgb]{0,0,0}{\makebox(0,0)[lb]{\smash{\emph{Figure 3: The extension of $f\in H^2$ off $Q$ is obstructed by $g\in H^3$.}}}}} 
  \end{picture}%
\endgroup
\end{center}

\subsection{Formal neighbourhoods}

We are interested in determining to what extent the cohomology classes $H^2(Q^I;\cO(k-2))$ and $H^3(Q^I;\cO(-k-4))$ can be extended off the quadric. The natural language in which to approach this question is that of formal neighbourhoods \cite{Pool,Witten:1978xx,Griffiths,Baston:1987av,Eastwood:1987}; a step by step process in which we consider the task of extending a class defined to $r$th order to the $(r+1)$th order about  $Q\subset\C\P^7$. The starting point is a twistor cohomology class  $f(Z)\in H^{\bullet}(Q^I;\cO(n))$ defined only on $Z^2=0$. The extension to the first formal neighbourhood is given by introducing the commuting variable $\xi$ such that $\xi^4=0$ and allowing the twistor function to now depend on $(Z^{\alpha},\xi)$, where $Z^2=\xi^2$. Similarly, we can think of the twistors associated with the $r^{\mathrm{th}}$ formal neighbourhood as the usual twistor $Z^\alpha$ with an additional variable $\xi$, for which $\xi^{2(r+1)}=0$, subject to
$$
(Z^{\alpha},\xi)\sim (tZ^{\alpha},t\xi)	\qquad	Z^2=\xi^2	\qquad	\xi^{2(r+1)}=0
$$
This leads to a \emph{thickening} of twistor space $Q\rightarrow Q_{[r]}$ \cite{Eastwood:1986} where $Q_{[r]}$ can be thought of as $Q$, but with the enlarged sheaf of holomorphic functions $\cO_{[r]}$ that contain the information of the first $r$-derivatives off $Q$ into $\CP^7$ encoded in the dependence on $Z$ and $\xi$.

More formally, we make use of the long exact sequence of cohomology groups that follows from the short exact sequence
$$
0\rightarrow \cO_{Q}(n-2r) \stackrel{\times (Z\cdot Z)^r}\longrightarrow
\cO_{[r]}(n) \longrightarrow \cO_{[r-1]}(n)\rightarrow 0\, .
$$
So $\cO_{[r]}$ is the sheaf of functions on the $r$th formal
neighbourhood which can be thought of as local functions on a neighbourhood of $Q$ in $\CP^7$ modulo
local functions of the form $(Z\cdot Z)^{r+1} \times g$ where $g$ is
another local function, with $\cO_{Q}=\cO_{[0]}$. In terms of the ideal sheaf $\mcI$ of $Q$, we have $\mcO_{[r]} = \sum_{i=0}^{r} \mcI^r / \mcI^{r+1}$.  In the following we
abbreviate $\cO_Q$ to $\cO$.

This short exact sequence leads to long exact sequences of cohomology groups in the usual way.  We know from \cite{Baston:1989vh} that the cohomology groups that are only non-zero at the $H^2(Q^I;{\cal O}(n))$ level for $n\geq -2$ and $H^3(Q^I;{\cal O}(n))$ for $n\leq -4$.   
To give an idea of the process we first consider the easy case of  extending $H^3(\cO(n))$ for $n\leq -4$.  The long exact sequence gives
$$
0\rightarrow H^3(\cO(n-2r)) \rightarrow H^3(\cO_{[r]}(n)) \rightarrow
H^3(\cO(n)) \stackrel{\delta}\rightarrow  0\, .
$$
The zeros arise because there are no $H^2$s in these homogeneity
degrees and there are never any $H^4$s.  Thus the obstruction
$\delta$ to extending a class from one formal neighbourhood to a
higher one always vanishes.  However, there is always the
freedom arising from adding the choice of an element of
$H^3(\cO(n-2r))$ to some original choice.  So, we can extend any class
in $H^3(\cO(n))$ to all orders but with much ambiguity, with a new
nontrivial term  arising at each term in its Taylor series.

The problem of extending $H^2(Q^I,\cO(n))$ for $n\geq -2$ is much less trivial.  For
the extension to the first formal neighbourhood $\cO_{[1]}$, the long exact sequence gives
$$
0\rightarrow H^2(\cO(n-2))\rightarrow H^2(\cO_{[1]}(n)) \rightarrow
H^2(\cO(n)) \stackrel{\delta}\rightarrow H^3(\cO(n-2)) \rightarrow \ldots .
$$

The case $n=-2$ is exceptional here as $H^3(\cO(n-2))=H^3(\cO(-4))$ is only non-vanishing
in this case and by the Penrose transform is given by solutions to the wave equation and so is indeed isomorphic to $H^2(\cO(-2))$.  In fact we will see by direct computation that $\delta$ is an isomorphism, so
that it is impossible to extend an element of $H^2(\cO(-2))$ to
$H^2(\cO_{[1]}(-2))$.  In fact this sequence then implies that
$H^2(\cO_{[1]}(-2))=0$ as $H^2(\cO(-4))=0$. 

For $n>-2$, $H^3(\cO(n))=0$ so that the
obstruction map $\delta$ always vanishes, and we can always extend
subject to an ambiguity in $H^2(\cO(n-2))$.  This ambiguity is trivial
for $n=-1$ and in that case there is a unique extension
$H^2(\cO_{[1]}(-1))=H^2(\cO(-1))$, but in all 
other cases there is some ambiguity.  Finally for $n\geq 0$, we now have
no obstruction to extension as the corresponding $H^3(\cO(n-2))$
vanishes, but we have an ambiguity of $H^2(\cO(n-2))$ in the choice of
extension.  

At the higher orders we obtain
$$
0\rightarrow H^2(\cO(n-2r))\rightarrow H^2(\cO_{[r]}(n)) \rightarrow
H^2(\cO_{[r-1]}(n)) \stackrel{\delta}\rightarrow H^3(\cO(n-2r))
\rightarrow \ldots . 
$$
For $n=-1$ and $r=2$ we now see that there is a possible obstruction
to extension as $H^2(\cO_{[1]}(-1))=H^2(\cO(-1))=H^3(\cO(-5))$ and
indeed, as we shall see later the map $\delta$ is again an
isomorphism. Thus $H^2(\cO_{[2]}(-1))=0$ and there is no extension to
the second formal neighbourhood in this homogeneity.

The other case we will be interested in is homogenieity $n=0$.  As we saw earlier, an $\alpha\in H^2(\cO(0))$ could always be extended to an $H^2(\cO_{[1]}(0))$ but with freedom in $H^2(\cO(-2))$.  Here at the next order $r=2$,  the obstruction group is now $H^3(\cO(-4))$ and indeed.  This map can be non-zero because, in $H^2(\cO_{[1]}(0))$ we had the summand (the freedom in extension to $H^2(\cO_{[1]})$) consisting of $H^2(\cO(-2))$ which is isomorphic to $H^3(\cO(-4))$. We will see again later that $\delta$ is an isomorphism from this summand and so we can always choose an extension of $\alpha$ to $\alpha_{[1]}\in H^1(\cO_{[1]}(0))$ so that $\delta\alpha_{[1]}=0$.   This condition fixes the ambiguity in the first step completely and there exists a unique $\alpha_{[2]}\in H^2(\cO_{[2]})$ that maps onto $\alpha_{[1]}$ because at this homogeneity, $H^2(\cO(-4))=0$.  
Now, at $r=3$ the obstruction to further extension is $\delta(\alpha_{[2]})\in H^3(\cO(-6))$.  This latter space is now isomorphic to the space $H^2(\cO)$ that we started with and in fact the map $\delta$ will be seen to be an isomorphism.   

\subsubsection{Explicit Example: Momentum Eigenstates}

Our task here is to explicitly compute the `connecting homomorphism' $\delta$ using the spinor helicity based twistor representatives and to show that it is an isomorphism whenever the cohomology groups it connects are related by the twistor transform.  In fact the action of  $\delta$ on Dolbeault representatives is clear.   If we take the first nontrivial case  $\delta: H^2(Q^I;\cO(-2)) \rightarrow H^3(Q^I;\cO(-4))$, with $\varphi\in H^2(Q^I;\cO((-2))$ we should, abusing notation, redefine $\varphi$ to be an arbitrary  smooth $(0,2)$-form on a \emph{neighbourhood} of $Z^2=0$ that agrees with the originally chosen $\varphi$.  Then, because the original $\varphi$ is a cohomology class, we will have $\dbar \varphi=0$ on $Z^2=0$, so on a neighbourhood we must have  
$$
\dbar \varphi=Z^2 \delta \varphi
$$
for some $(0,3)$-form $\delta \varphi$.  If we change the extension of $\varphi$ to $\varphi+ Z^2 \Lambda$, then $\delta \varphi\rightarrow \delta \varphi + \dbar \Lambda|_{Z^2=0}$ so that the cohomology class of $\delta \varphi$ is well-defined.

We first show that for $\varphi\in  H^2(\cO(-2))$, $\delta \phi \in H^3(\cO(-4))$ is precisely the twistor transform of the class that we started with corresponding to the same solution to the wave equation $\Phi$ on space-time.  
We have that 
$$
\varphi= \int_{\C\P^1} \D u \, \, \bar\delta^3_{-2,-2}(\pi,\lambda\cdot u)\,\e^{- \ii \omega\cdot \widehat\lambda}  ,\qquad 
\phi= \int_{\C\P^1} \D u \, \D \hat u\, \bar\delta^3_{-4,0}(\pi,\lambda\cdot u)\,\e^{- \ii \omega\cdot \widehat\lambda} \, .
$$
The classes  clearly extend off $Z^2=0$ and, using (\ref{dbar-closure}) we see that $\phi$ is $\dbar$-closed.  There is clearly much freedom in extending these representatives off the quadric as we can add on $Z^2$ multiplied by any form of the same degree and with homogeneity two lower; however, the form $\phi$ is not closed away from $Z^2=0$.   Following the calculation in \eqref{dbar-closure} but not imposing the condition $Z^2=0$ we find
$$
\dbar \varphi = - \ii\omega\cdot\pi \int_{\CP^1} \D u\, \D\hat u \,  \e^{- \ii \omega\cdot \widehat\lambda} \,
\bar\delta^3(\pi,\lambda)  = - \ii Z^2 \phi 
$$
Thus $\delta\varphi= - \ii\phi$ and this gives the non-triviality of the map $\delta$.

Similar calculations can be done with other homogeneities.  We shall be particularly interested in twistor representatives  corresponding to $k=0,1$ and $2$
and so we only deal with these cases here. 

For $k=1$ we have  representatives $\chi\in H^2(Q^I,\cO(-1))$ and $\psi\in H^3(Q^I,\cO(-5))$ corresponding to momentum eigenstates of the spin-half field $\Psi_A(x)$ given by
$$
\chi= \int_{\C\P^1} \D u  \, \xi\cdot u \,
\bar\delta^3_{-1,-3}(\pi,\lambda)\, \e^{- \ii \omega\cdot \widehat\lambda}\, ,\qquad 
 \psi= \int_{\C\P^1} \D u \, 
{\D \hat u}\,\xi\cdot \hat u \,  \, \bar\delta^3_{-5,1}(\pi,\lambda)\,\e^{- \ii \omega\cdot \widehat\lambda}\, .
$$
As before we calculate the anti-holomorphic derivative of the $H^2$ representative
$$
\dbar \chi =  - \ii \omega\cdot \pi\,\int_{\CP^1} 
\D u \,   \D\hat u \, \xi\cdot u
 \, \bar\delta^3(\pi,\lambda)\, \e^{- \ii \omega\cdot \widehat\lambda}
 \, . 
$$
All this does not vanish to first order in $Z^2=\omega\cdot\pi$.
However, to first order in $Z^2$ we must have that this is exact by
the abstract  arguments above.  To see this, we observe that  
$$
\dbar \int_{\CP^1} 
\D u \, \xi\cdot \hat u \,
\bar\delta^3(\pi,\lambda)\,  \e^{- \ii \omega\cdot \widehat\lambda}=\int_{\CP^1} 
\D u \, {\D\hat  u} \, (\xi\cdot u)
\, \bar\delta^3(\pi,\lambda)\e^{- \ii \omega\cdot \widehat\lambda} 
- \ii \omega\cdot \pi \psi \, . 
$$
with the first term arising from $\dbar \xi\cdot\hat u$ and the second from $\dbar$ of the exponential. Thus, if we redefine 
$$
\chi= \int_{\C\P^1} \D u  \, \left (\xi\cdot u \, \bar\delta^3_{-1,-3}(\pi,\lambda) + \ii (\xi\cdot \hat u)\, ( \omega\cdot \pi )\, \bar  \delta^3_{-3,-1}(\pi,\lambda)\right)\, \e^{- \ii \omega\cdot \widehat\lambda}
$$
we obtain the desired relation
$$
\dbar \chi= - \ii Z^4\psi \, .
$$

The final case of interest to us is the case corresponding to spin-one fields, described in twistor space by $b\in H^2(Q^I,\cO)$ and $h\in H^3(Q^I,\cO(-6))$.  Following the above strategy, we now see that if we redefine 
\begin{multline*}
b=\int_{\C\P^1} \D u \, \e^{- \ii \omega\cdot \hat\lambda} \, \left (\xi_{ab} u^au^b
  \, \bar\delta^3_{0,-4}(\pi,\lambda) - {\xi_{ab}u^a \hat
    u^b}\,  \omega\cdot \pi \, \bar
  \delta^3_{-2,-2}(\pi,\lambda) \right. \\
\left. + \xi_{ab}\hat u^a\hat
    u^b (\omega\cdot \pi)^2\bar \delta^3_{-4,0}
  (\pi, \lambda)\right) 
\end{multline*}
and 
$$
 h = \int_{\C\P^1} \D u \, \D \hat u\, \xi_{ab} \hat u^a\hat u^b \, \bar\delta^3_{-6,2}(\pi,\lambda) \,
\e^{- \ii \omega\cdot \widehat\lambda}\, ,
$$
then we will have 
\begin{equation}\label{gerbefn}
\dbar b =Z^6h
\end{equation}
as desired.  By the abstract arguments of the previous subsection, the
final forms of these representatives are unique. In general we have
\be{fn}
\dbar f=Z^{2k+2}\,g
\ee
where $g(Z)\in H^3(Q^I;(-k-4))$ and $f(Z)\in H^2(Q^I;(k-2))$.

\subsection{Integral formulae in the $k-2$ homogeneity case}

The above uniquely extended representatives in the $k-2$ homogeneity case,
together with their connection to the negative homogeneity case
allow us to define integral formulae. Recalling the form of the integral expression for the direct Penrose transform
$$
F_{A_1\ldots A_k}(x)=\int_{S_x}\D^3\pi\;\pi_{A_1}\ldots
  \pi_{A_k}\, g(\omega,\pi) \, .
$$
where  $g(Z)\in H^3(Q^I;(-k-4))$. Combing this with the expression (\ref{fn}), we have 
\be{indirect-h2a}
F_{A_1\ldots A_k}(x)=\int_{S_x}\D^3\,\pi Z^{-2k-2}\,\pi_{A_1}\ldots
  \pi_{A_k}\,\dbar f(\omega,\pi) \, .
\ee
where $f(Z)\in H^2(Q^I;(k-2))$. This integral formula hides the cumbersome fact that we have to construct the the canonical extension of $f$ off $Q$ so its practical use may well be rather limited.

\section{$\Xi$ and half-Fourier transforms in split signature}

As described in Section two, in split signature, the components of the twistor $Z^\alpha$ can be taken to be real and twistor space to be the real quadric of signature $(4,4)$ inside $\R\P^7$,  $Q=(S^3\times S^3)/\Z_2$. In this signature, the integral formula for solutions to the massless fields equations 
\be{direct-split}
F(x)_{A_1\ldots A_k}
=\int_{S_x} \D^3\pi\,
\pi_{A_1}\ldots\pi_{A_k}\,f\left(x^{AB}\pi_B,\pi_A\right) \, ,
\ee
can be taken as an integral now over $S_x=\RP^3$ with $f$ being a straightforward smooth function on $Q$ of weight $-4-k$ rather than as a representative of some $H^3$ cohomology class.  Thus we have the benefit in split-signature that \v Cech
and Dolbeault  representatives are replaced by real functions on the real slice where all twistor coordinates and space-time coordinates are real. Tree amplitudes generally are rational functions and so extend to split signature real slices.  This real approach simplifies matters significantly and allows one to exploit standard tools such as Fourier analysis.
Sparling \cite{Sparling:2006zy,Sparling:2007} referred to the $k=0$ version of this transform as the $\Xi$-transform, and we will follow his terminology here.  We will use Fourier analysis to identify the kernel of this map on twistor space and to extend it to the `indirect' cases of weight $k-2$).

In four dimensions with split signature the Penrose transform similarly has a non-cohomological analogue, the X-ray transform, that maps functions on the real twistor space, $\RP^3$ to solutions to the massless field equations on space-time by straightforward integration along lines in the real twistor space.   This can be combined with the Fourier transform to give a
map from functions on twistor space to functions on the light-cone in momentum space.  This yields what has
become known as the `half-Fourier transform' \cite{Witten:2003nn}. Both the momentum light cone and twistor space are three-dimensional and the map is a Fourier transform along a natural family of two-dimensional fibres.   

In this section we derive the analogue of this construction for six-dimensional space-time with split signature.
However, the real momentum space light-cone is 5-dimensional whereas the real twistor space is 6-dimensional and we will find that this leads to new features in the correspondence as it can no longer be one to one.  In this context, the transforms have been studied by Sparling \cite{Sparling:2006zy,Sparling:2007} for the wave equation and homogeneities $-2$ and $-4$ who referred to them as the $\Xi$-transform.    He discovered that the $-4$ case leads to solutions to the wave equation but the kernel of the map consists of those twistor functions $F$ that are in the image of the conformally invariant wave operator $\Box$ on {\em twistor space}, $f=\Box g$ for some $g$  (recall that twistor space, being a quadric, is canonically a conformal manifold in six dimensions).  Using triality he was able to follow this around the correspondences between twistor space, primed twistor space and space-time with them all on an equal footing.  Here we will extend this to all other weights, both positive and negative and see how the conformally invariant powers of the Laplacian of \cite{EastwoodRice,GJMS} play a role in characterizing the twistor data in this case.

Twistor space is a conformal manifold and so admits conformally invariant powers of the ultrahyperbolic wave operator \cite{GJMS}
$$
\Box^{k+1}:\mcE(-k-2)\rightarrow \mcE(-k-4)\, .
$$
Here, $\mcE(k)$ denotes the sheaf of smooth sections over $\mcQ$ homogeneous of degree $k$.

We will show 
\begin{propn}
The kernel of the $\Xi$-transform (\ref{direct-split}) for functions on twistor space of weight $-k-4$ is the image of  
$$
\Box^{k+1}: \Gamma ( Q , \mcE(-k-2) ) \rightarrow \Gamma ( Q, \mcE(-k-4) )\, .
$$
The $\Xi$-transform therefore gives an isomorphism 
$$
\Gamma_k(\M)\simeq \Gamma(Q,\mcE(-k-4))/\{ \mbox{ Im } \Box^{k+1}:\Gamma ( Q , \mcE(-k-2) ) \rightarrow \Gamma ( Q, \mcE(-k-4) )\} 
$$
\end{propn}
and as the analogue of the indirect Penrose transform we will show 
\begin{propn}
For $k>0$,there is a one to one correspondence 
$$
\{h\in \Gamma(Q,\mcE(k))| \Box^{k+1} h=0\}\simeq \Gamma_k(\M)\, .
$$
\end{propn}
Our main tool will be the six-dimensional analogue of the half-Fourier transform.  However, before we embark on that, we remark that the spinor-helicity representatives for cohomology classes that we obtained earlier  have totally real analogues that we write down directly here as
\begin{eqnarray}
\phi(\omega,\pi)=\int_{\R\times\RP^1_u} \e^{- \ii s \omega\cdot\widehat\lambda}
\, \, \alpha_\xi 
\delta^4(s\pi-\lambda\cdot u) \,
s^{k+3}\rd s \wedge \D u \, ,  \label{h3-spin-hel-rep-real}
\\
\psi(\omega,\pi)=\int_{\R\times \RP^1_u} \e^{- \ii s\omega\cdot\widehat\lambda}  \, \, \xi( u) \, 
\delta^4(s\pi-\lambda\cdot u)   \, \frac{\rd s}{s^{k-1}}\wedge \D u \, . \label{h2-spin-hel-rep-real}
\end{eqnarray} 
where now $Z, \lambda_A^a, u_a$ and $s$ are real, $\hat u_a=(u_1,-u_0)/(u_0^2 +u_1^2)$ as before but is real and $\alpha_\xi$ is a smooth function of the $u_a$ of homogeneity $-k-2$ satisfying
$$
\int_{\RP^1_u}u_{a_1}\ldots u_{a_k} \alpha_\xi \D u=\xi_{a_1\ldots a_k}\, .
$$
Here, $\alpha_\xi$ can still be thought of as a representative of a class in $H^1(\CP^1,\cO(-k-2)$, but now as a \v Cech cocycle defined on $\RP^1\subset \CP^1$.   It is easily seen that substituting \eqref{h3-spin-hel-rep-real} into \eqref{direct-split} gives the spinor-helicity momentum eigenstate as expected.

Analogues of these representatives can be found in higher dimensions also as before.

\subsection{The Fourier and half-Fourier transform}

We first formulate the Fourier transform from space-time for a field
$F(x)_{A_1\ldots A_k}$.  Since it satisfies the massless field equation
$\nabla^{AB}F_{B\ldots D}=0$, the transform to momentum space 
$$
\widetilde{F}(P)_{A_1\ldots A_k}=\int\rd^6x\;F(x)_{A_1\ldots
  A_k}\;\e^{\ii P\cdot x} 
$$
will
satisfy $P^{AB}\widetilde{F}_{B\ldots D}=0$.  Thus $\widetilde{F}$ only has support when $P^{AB}$ has rank two. In particular  $P$ is null and $F$ is
supported on the momentum light cone 
$$
M_0=\{P^{AB}|P^2=0\}=\{P_{AB}|P_{AB}=\varepsilon_{ab}\lambda^a_A\lambda^b_B\} 
$$
For some $\lambda^a_A$ defined up to $\SL(2,\R)$ on the $a$ index. As before, we obtain the spinor helicity representation
$$
\widetilde{F}_{A_1\ldots A_k}=\widetilde{F}(\lambda)_{a_1\ldots  a_k}
\lambda^{a_1}_{A_1} \ldots \lambda^{a_k}_{A_k}\delta(P^2),
$$ 
and define $\widetilde{F}_{a_1\ldots  a_k}$ to be the Fourier transform of the field $F(x)_{A_1\ldots A_k}$.  

Our next task is to establish the following.

\begin{propn}
We have the following direct formula for $\widetilde{F}_{a_1\ldots
  a_k}$ in terms of the twistor function $f$ that gives rise to it via
\eqref{direct-split} in split signature 
\be{fourier1}
\widetilde{F}(\lambda^a)_{a_1\ldots a_k}=\int \D u\;  u_{a_1}\ldots
u_{a_k}\int\rd^4\omega\;\delta(\omega^A\lambda_A)\,f(\omega^A,\lambda_A)\;\e^{-\ii \widehat{\lambda}_A\omega^A}\,
. 
\ee
where
$
\lambda_A=\lambda^a_Au_a$ where $u^a$ are now homogeneous coordinates on $\RP^1$ and $\widehat{\lambda}_A$ is any spinor
chosen so that $\lambda_{[A}\widehat\lambda_{B]}=P_{AB}$.
\end{propn}
For definiteness we will often choose $\widehat
\lambda_A=\lambda^a_A\hat{u}_a$ 
where $\hat u_a=(-u_1,u_0)/(u_0^2+u_1^2)$ but our formulae will be
invariant under $\widehat\lambda\rightarrow \widehat\lambda
+\alpha\lambda$ for all $\alpha$.

\medskip

\noindent
{\bf Proof:} In split signature all quantities, $P$, $x$, $\lambda$, $u$, $\hat u$,
$\omega$ and $\pi$ can be
taken real and, at least for homogeneity $-k-4$, the twistor
cohomology class can be replaced  by a smooth function $f$ of the real twistor variables $(\omega^A,\pi_A)$ of projective weight $-k-4$. The direct Penrose
transform gives a  space-time massless field
$$
F(x)_{A_1\ldots A_k}
=\int_{S_x} \D^3\pi\,
\pi_{A_1}\ldots\pi_{A_k}\,f\left(x^{AB}\pi_B,\pi_A\right) \, ,
$$
where now the integral is over $S_x=\RP^3$.
We can Fourier transform this to get a function on momentum space and
substituting in its above form, we must obtain
$$
\widetilde{F}(\lambda)_{a_1\ldots a_k}\lambda^{a_1}_{A_1} \ldots
\lambda^{a_k}_{A_k}\delta(P^2)=\int\rd^6x\; \D^3\pi\;\pi_{A_1}\ldots \pi_{A_k}
\, f\left(x^{AB}\pi_B,\pi_A\right) \;\e^{\ii P\cdot x}\, . 
$$

We now reverse the order of integration performing the $x$-integral first.  We can reparametrise $x$ with $\omega^A$ together with a parameter $\chi_A$ defined up to the addition of multiples of $\pi$ by
$$
x_{AB}=\varepsilon_{ABCD}\omega^C\alpha^D+\chi_{[A}\pi_{B]}\, .
$$
We can choose $\alpha^A$ so that $\alpha\cdot \pi=1$ and fix the freedom in $\chi$  so
that $\alpha\cdot \chi =0$.  This is then consistent with the
incidence relation $\omega^A=x^{AB}\pi_B$ and $\chi_A$ has three
independent components (and projective weight -1). 
We may now write
\begin{eqnarray}
P\cdot x&=&2P_{AB}\omega^A\alpha^B+P^{AB}\chi_A\pi_B\nonumber
\end{eqnarray}
The measure on space-time can then be written as
$$
\rd^6x=\rd^4\omega\;\D^3\chi\;\delta(\omega^A\pi_A)\; 
$$
where
$\D^3\chi=\varepsilon^{ABCD}\pi_A\;\rd\chi_B\;\rd\chi_C\;\rd\chi_D(=\rd^4\chi
\;\delta(\alpha^A\chi_A))$. Thus
we have 
$$
\widetilde{F}(\lambda)_{A_1\ldots
  A_k}\delta(P^2)=\int\rd^4\omega\;\D^3\chi\;\D^3\pi\;\delta(\omega\cdot\pi)\;
\pi_{A_1}\ldots\pi_{A_k}\,f(\omega^A,\pi_A)\;\e^{2\ii P_{AB}
  \omega^A\alpha^B + \ii P^{AB}\chi_A\pi_B}  
$$
Performing the $\chi$ integration we have\footnote{To see this, choose
  $\alpha$ so that
  $\alpha\cdot \chi=\chi_1=0$ 
  so that
$\int D^3\chi\;  \e^{\ii \chi_IP^{IA}\pi_A}=\prod_{I=2,3,4}\delta^3(P^{IA}\pi_A).
$
On $P_{12}\neq 0$, we can choose $\lambda^0_1=1$
$\lambda^0_2=\lambda^1_1=0$ and $\lambda^1_2=P_{12}$ and we find 
  $$
|P_{12}|\delta(P^{3A}\pi_A)
  \delta(P^{4A}\pi_A)=\delta^2(\pi_A,\lambda_A^a).$$ Consider  now
  $\delta(P^2)\delta^2(\pi_A;\lambda^+,\lambda^-)=|P_{12}|\delta(P^{12}P^{34}+P^{14}P^{23}+P^{13}P^{42})\delta(P^{3A}\pi_A)\delta(P^{4A}\pi_A)$. On
  the support of the second and third delta-functions, the first can
  be rewritten and we find
  $\delta(P^2)\delta^2(\pi_A;\lambda^a)=\delta^3(P^{IA}\pi_A)
  $ and therefore $\int
  \D^3\chi\;\e^{\ii P^{AB}\chi_A\pi_B}=\delta(P^2)\delta^2(\pi_A,\lambda^a_A)$.}
$$
\int
\D^3\chi\;\e^{\ii P^{AB}\chi_A\pi_B}=\delta(P^2)\;\int_{\R\P^1}\D u\;\delta^3_{-2}(\pi_A,\lambda_A^au_a)\, .
$$
On the support of this $\delta$-function all the $\pi_A$ factors can be replaced by $\lambda^a_Au_a$ and we can now remove the $\lambda_A^a$ and  $\delta(P^2)$ factors from both sides of the equation to obtain
$$
\widetilde{F}(\lambda)_{a_1\ldots a_k}=\int \D u \; u_{a_1}\ldots u_{a_k}  \int\rd^4\omega\;\D^3\pi\; \delta(\omega^A\pi_A)\;\delta^3(\pi_A;\lambda_A\cdot u)\;  f(\omega^A,\pi_A)\;\e^{2\ii P_{AB}\omega^A\alpha^B}
$$
Using
$\varepsilon_{ab}=(u_a\hat{u}_b-u_b\hat{u}_a)$,
on the support of the delta-functions, the exponent may be written 
$$
2\ii P_{AB}\omega^A\alpha^B=-\ii\widehat{\lambda}_A\omega^A
$$
where
$
\lambda_A=\lambda^a_Au_a$ and $	\widehat{\lambda}_A= 2\lambda^a_A\hat{u}_a
$.
With this, and using the $\delta^3(\pi_A,\lambda_A)$ to do the $\D^3\pi$ integration, the map from functions on twistor space to scalar fields on the null cone becomes
\be{fourier2}
\widetilde{F}(\lambda^a)_{a_1\ldots a_k}=\int \D u\;  u_{a_1}\ldots u_{a_k}\int\rd^4\omega\;\delta(\omega^A\lambda_A)\,f(\omega^A,\lambda_A)\;\e^{-\ii \widehat{\lambda}_A\omega^A}\, .
\ee
as required.$\Box$

\begin{center}
\begingroup
  \makeatletter
  \providecommand\color[2][]{%
    \errmessage{(Inkscape) Color is used for the text in Inkscape, but the package 'color.sty' is not loaded}
    \renewcommand\color[2][]{}%
  }
  \providecommand\transparent[1]{%
    \errmessage{(Inkscape) Transparency is used (non-zero) for the text in Inkscape, but the package 'transparent.sty' is not loaded}
    \renewcommand\transparent[1]{}%
  }
  \providecommand\rotatebox[2]{#2}
  \ifx\svgwidth\undefined
    \setlength{\unitlength}{12cm}
  \else
    \setlength{\unitlength}{\svgwidth}
  \fi
  \global\let\svgwidth\undefined
  \makeatother
  \begin{picture}(1,0.70707072)%
    \put(0,0){\includegraphics[width=\unitlength]{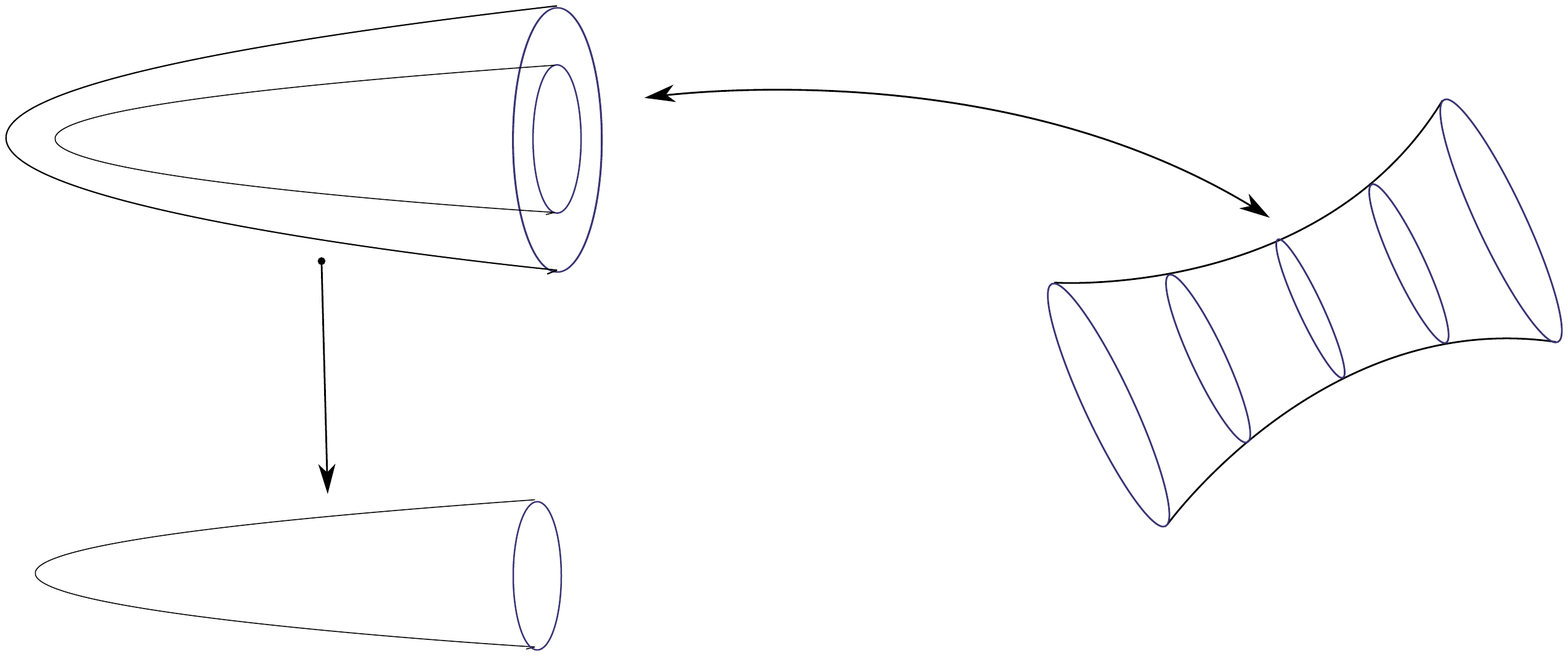}}%
    \put(0.12426258,0.5756281){\color[rgb]{0,0,0}\makebox(0,0)[lb]{\smash{$T(-2)\R\P^3$}}}%
    \put(0.16994733,0.15350081){\color[rgb]{0,0,0}\makebox(0,0)[lb]{\smash{$M_0$}}}%
    \put(0.67796198,0.23390601){\color[rgb]{0,0,0}\makebox(0,0)[lb]{\smash{$Q$}}}%
    \put(0.28324558,0.34354945){\color[rgb]{0,0,0}\makebox(0,0)[lb]{\smash{$\int \D u$}}}%
    \put(0.551872,0.51532419){\color[rgb]{0,0,0}\makebox(0,0)[lb]{\smash{Half Fourier Transform}}}%
\put(0.0,0.0){\color[rgb]{0,0,0}{\makebox(0,0)[lb]{\smash{\emph{Figure 4: The half Fourier transform relates a bundle over $M_0$ to $Q$.}}}}} 
  \end{picture}%
\endgroup
\end{center}

\medskip
This map factors into a {\em half Fourier transform} followed by an
integral over $u$.  
The half-Fourier transform part of the map, being a
Fourier transform, is necessarily one-to-one.  This takes functions on
twistor space $Q$ to functions on an extension of the momentum
light-cone $M_0$  by $\RP^1$ to include the real projective coordinate $u^a$
giving a six-dimensional auxiliary space with coordinates
$\{\lambda_A^a,u^a\}/\{\SL(2,\R)\times \R^*\}$ with the $\SL(2,\R)$ acting on
the $a$-index and the $\R^*$ on $u^a$.  We identify this space 
with the bundle $T(-2)\RP^3$ of tangent vectors 
of homogeneity $(-2)$ over $\RP^3$.  This follows by rewriting the data
$(\lambda^a_A,u^a)$ as 
$$
\left(\lambda_A,\widehat\lambda_B\right):=\left(\lambda\cdot u,\lambda\cdot\hat u\right)\longrightarrow  \left(\lambda_A\cdot u,\lambda_A\cdot\hat
u\frac{\p}{\p\lambda_A} \right)\in
\RP^3 \times T_{\lambda}\RP^3
$$
$\widehat \lambda$ is only defined up to the addition of multiples of
$\lambda$ but $\lambda\cdot\p/\p\lambda$ is zero in $T_\lambda
\RP^3$. 
 \begin{defn}
The half-Fourier transform takes a homogeneous function
 $f(\omega,\pi)$ of weight $-4-k$ on twistor space to a homogeneous
 function $K(\lambda,\widehat\lambda)$ on $T(-2)\RP^3$ of weight $2-k$ by
\be{}
K(\lambda,\hat\lambda)=\int\rd^4\omega\;\delta(\omega^A\lambda_A)\,f(\omega^A,\lambda_A)\;\e^{-\ii \widehat{\lambda}_A\omega^A}\, . 
\ee
We remark that in this definition, $k$ can be any integer.
\end{defn}

 The delta function in the integrand implies that
 $K(\lambda,\widehat\lambda+\alpha\lambda)=K(\lambda,\widehat\lambda)$
 for all $\alpha$.
This gives an analogue of the half Fourier transform familiar from four-dimensions. It is invertible with inverse
$$
f(\omega^A,\lambda_A)\;\delta(\omega^A\lambda_A)=\frac 1{(2\pi)^4}\int\rd^4\widehat{\lambda}\;  K({\lambda}_A,\widehat\lambda_A)\;\e^{\ii \widehat{\lambda}_A\omega^A}
$$
where we think of $f(\omega^A,\lambda_A)\delta(\omega^A\lambda_A)$ as a distribution on $\RP^7$ with support on the twistor quadric, the delta function arising because $K(\lambda,\widehat\lambda)=K(\lambda,\widehat \lambda + r\lambda)$ for all $r$.

\subsection{The extended  direct $\Xi$-transform and half-Fourier transform}
  
Our proposition above therefore gives the Fourier transform of the massless field as
the following integral over $\RP^1$ of the half-Fourier transform $K$ of $f$
\be{int-rp1}
\widetilde{F}(\lambda^a_A)_{a_1\ldots a_k}=\int_{\RP^1} \D u\; u_{a_1}\ldots u_{a_k}K({\lambda},\widehat\lambda)\, .
\ee
The combination cannot be one-to-one since the dimension of twistor space is six and the dimension of the momentum null-cone
is five.  There must therefore be a kernel of the map \eqref{direct-split} from twistor functions to massless fields which we now seek to
identify. Clearly, since the half-Fourier transform is 1:1, the kernel is that of the integration in \eqref{int-rp1}, i.e., those $K$ for which the integrand is exact.  We will identify this and its counterpart on twistor space.  We will also in the process see how to extend this correspondence to the case of homogeneity $k-2$ which will work dually.

It is clear that the integration \eqref{int-rp1} has as kernel those $K$ for which the integrand is exact for the $\RP^1$ integration.    In order to understand this kernel, we need to review the theory of the $\eth^k$ operators on $\RP^1$.  These are operators defined in the first instance on functions on $\RP^1$, for $k \geq 0$,
$$
\eth^{k+1}:\mcE(k)\rightarrow \mcE(-k-2)
$$
that are invariant under Mobius transformations.  Following \cite{Eastwood:1981} we observe that for $f(u)$ of weight $k$,
$\frac{\p^kf}{\p u^{a_1}\ldots \p u^{a_k}}$ has weight 0 so that 
$$
u^{a_0}\frac{\p^k f}{\p u^{a_0}\ldots \p u^{a_k}} =0\, . 
$$
Thus
$$
\frac{\p^k f}{\p u^{a_0}\ldots \p u^{a_k}}=u_{a_0}\ldots u_{a_k}  \eth^{k+1}f
$$
for some $\eth^{k+1}f$ of weight $-k-2$.  This defining equation can be written
\be{d-eth}
\rd \left(\frac{\p^k f}{\p u^{a_1}\ldots \p u^{a_k}}\right)=u_{a_1}\ldots u_{a_k}  \eth^{k+1}f \, \D u
\ee
thus showing that $K$ is in the kernel of the integral \eqref{int-rp1} iff $K=\eth^{k+1}J$ for some $J$.

This gives  a complete characterization of the kernel of the integration in  \eqref{int-rp1} where now $\eth^{k+1}$ is understood fibrewise  for the fibres of the fibration $T(-2)\RP^3\rightarrow M_0$.   In order to understand what this means on twistor space, we first express $\eth$ in terms of the $(\lambda,\widehat \lambda)$ coordinates on $T(-2)\RP^3$.  We first observe that if we are not concerned to preserve manifest projective invariance on $\RP^1$, then we can define $\eth^{k+1}=(\eth)^{k+1}$ where on a function $f$ of weight $k$ we define the action of a single $\eth:\mcE(k)\rightarrow \mcE(k-2)$ for all integral values of $k$ by
$$
\eth f=(u_0^2+u^2_1)^{k/2} \eth\left( (u_0^2+u_1^2)^{-k/2} f\right)\, .
$$
It is now possible to see that, with this definition, $(\eth)^{k+1}$ is independent of the choice of  quadratic form (here $(u_0^2+u_1^2)$) and reduces to the invariant form given above.  As an operator on $K(\lambda,\widehat\lambda)$ with $K(\lambda,\widehat\lambda +\alpha\lambda)$ we find, using the chain rule, together with $\eth u_a=\hat u_a$ and $\eth \hat u_a=u_a$, that
$$
\eth K(\lambda,\widehat\lambda)= \widehat \lambda\cdot\frac{\p K}{\p\lambda}\, .
$$
It is now straightforward to see that, since under the half-Fourier transform, $\widehat\lambda \leftrightarrow \p/\p\omega^A$,  we have
$$
\eth K\leftrightarrow \frac{\p^2 f}{\p \omega^A\p\pi_A}\, .
$$
Thus $\eth$ corresponds to the ultrahyperbolic wave operator $\Box$ on the non-projective $\R^8$ associated to the $\RP^7$ in which $Q$ lives.  Homogeneous functions on $\RP^7$ do simply correspond to ordinary functions on this $\R^8$, but this is in general not well defined on functions on $Q$ as the  derivative off the quadric $Q$ is not given.  However, for weights $k\geq 0$ the following operators are well defined on homogeneous functions on $Q$
$$
\Box^{k+1}:\mcE(k-2)\rightarrow \mcE(-k-4)
$$
as defined in precisely this context in \cite{GJMS} (see also \cite{EastwoodRice, Hughston:1984gz,FeffermanGraham} for antecedents).
Thus we have 
\begin{propn}
The kernel of the $\Xi$-transform (\ref{int-rp1}) for functions on twistor space of weight $-k-4$ is the image of  
$$
\Box^{k+1}:\Gamma ( Q , \mcE(-k-2) ) \rightarrow \Gamma ( Q , \mcE(-k-4) )\, .
$$
The $\Xi$-transform therefore gives an isomorphism 
$$
\Gamma_k(\M)\simeq \Gamma(Q,\mcE(-k-4))/\{ \mbox{ Im } \Box^{k+1}:\Gamma ( Q , \mcE(-k-2) ) \rightarrow \Gamma ( Q , \mcE(-k-4) )\} \, .
$$
\end{propn}

\subsection{The indirect $\Xi$-transform}

The question remains as to how the $\Xi$-transform works for
homogeneity $k-2$.  The half-Fourier transform is an isomorphism from
functions $h$ of weight $k-2$ on twistor space to functions $J$ of
weight $k$ on $T(-2)\RP^3$.    The map from a momentum space
representative $\widetilde{F}(\lambda^a_A)_{a_1\ldots a_k}$ to such a
function $J$ of weight $k$ is clear: it should be
$$
J=\widetilde{F}(\lambda^a_A)_{a_1\ldots a_k}u^{a_1}\ldots u^{a_k}\, .
$$
This is inverted by setting
$$
\widetilde{F}(\lambda^a_A)_{a_1\ldots a_k}=\frac{\p^k}{\p
  u^{a_1}\ldots \p u^{a_k}}J(\lambda,\widehat \lambda)
$$
but if $J$ had been chosen arbitrarily, this will only give a sensible momentum space representative if
the right hand side is independent of $u_a$.  Using \eqref{d-eth} we
see that this will follow iff $\eth^{k+1}J=0$.  
Using the above argument, we can see that this is equivalent for the
function $h$ on 
twistor space to the
vanishing of the conformally invariant power of the ultrahyperbolic
wave equation $\Box^{k+1}h=0$.  Thus we have the positive chirality
$\Xi$-transform:
\begin{propn}
For $k>0$,there is a one to one correspondence 
$$
\Gamma_k(\M)
\simeq \{h\in \Gamma(Q,\mcE(k))| \Box^{k+1} h=0\}\, .
$$
\end{propn}
Thus the $\Xi$-transform maps solutions of differential equations on
one space to solutions on another in this case.  We remark also that
in split signature we have triality, so that twistor space,
primed twistor space and (compactified) space-time are all on an
equivalent footing and the $\Xi$-transform applies between any two of these
three spaces in either direction.  

We finally note the integral formula in this case
\begin{equation}\label{int-h2-real}
\phi_{A_1\ldots A_k}(x)=\int_{\gamma\subset S_x}  \frac{\p^{k+1} h}{\p\omega^{A_1}\ldots \p\omega^{A_{k}} \p\omega^{A_{k+1}}} \varepsilon^{A_{k+1}BCD}\pi_B\rd\pi_C\rd\pi_D\, ,
\end{equation}
where the integral is over some 2-dimensional contour $\gamma$ cohomologous to $\RP^2\subset S_x=\RP^3$.  In order to make sense of the $\p/\p\omega^A$ derivatives, $h$ must be extended off $Q$ to $k$th order in $\RP^7$; the $k+1$th derivative is skewed with $\pi_A$ and so is acting only tangent to $Q$.  This is precisely what is done in \cite{GJMS} with the extension determined by the condition that $\Box_{\R^8} h=0$ in the ambient non-projective space to the appropriate order;  the operator $\Box^{k+1}_Q$ on $h$ is obtained in \cite{GJMS} as the obstruction to extending $h$ as a solution to $\Box_{\R^8} h=0$ at $k+1$th order although there is in any case some ambiguity at that order.  The result is independent of the chosen contour by virtue of  $\Box^{k+1}h=0$ which in particular implies that the $k$th order extension is annihilated by $\Box_{\R^8}$.  We remark that it is not sufficient to check this with the spinor-helicity representative \eqref{h2-spin-hel-rep-real} except for the $k=0$ case as for general $k$ \eqref{h2-spin-hel-rep-real} has not been extended appropriately off $Q$ into $\RP^7$ and formulae more analogous to those used in the formal neighbouhood discussion  are required.

\section{Interactions and Scattering amplitudes}

There has been considerable work on the spinor-helicity construction of scattering amplitudes for (1,1) super Yang-Mills in six-dimensions \cite{Cheung:2009dc,Brandhuber:2010mm,Dennen:2010dh,Bern:2010qa,Bern:2010yg,Dennen:2009vk}. The invariances of the theory appear to uniquely fix the form of the three photon amplitude and higher point amplitudes have been derived using the BCFW construction \cite{Cheung:2009dc}\footnote{Loop amplitudes have also been constructed by unitarity methods in \cite{Brandhuber:2010mm,Bern:2010qa}.}. Despite a clear demonstration that spinor-helicity methods are useful in six dimensions, there is little understanding of how to construct scattering amplitudes from six-dimensional twistor space. We shall return to the question of how to describe non-chiral theories such as Yang-Mills in future work. In this article we are chiefly concerned with conformally invariant and chiral theories.  Because the amplitudes for the $(0,2)$-theory are believed to be trivial, we are rather limited as to the amplitudes we can consider and we focus on the $\Phi^3$ vertex.
(The only fields of spin less than one we can consider are scalars and spin-half fermions and it does not seem possible to construct an interaction involving spin-half fields without introducing a spin-half field $\Psi^A(x)$ of opposite chirality.)

We can write down a classically conformally-invariant Lagrangian for an interacting scalar field
\begin{equation}\label{scallag}
{\cal L}=\frac{1}{2}\partial_{\mu}\Phi\,\partial^{\mu}\Phi+\frac{\kappa}{6}\;\Phi^3 
\end{equation}
where $\kappa$ is a dimensionless coupling constant. 
By Fourier transform, the three-scalar amplitude corresponding to the $\Phi^3$ vertex takes the simple form
$$
{\cal A}(P_1,P_2,P_3)=\frac\kappa6\;\delta^6\left(\sum_{i =1}^3P_{AB}^{i }\right) \, .
$$

A natural  candidate for the three-point scalar amplitude in terms of indirect Penrose transform functions $\varphi_i $, $i =1,2,3$ of weight $-2$ is
\begin{equation}\label{amplitude}
{\cal A}(P_1,P_2,P_3)=\int_{\R\P^7} \D^7Z\;\delta(Z\cdot Z)\;\varphi_1\,\varphi_2\,\varphi_3\nonumber\, .
\end{equation}
We insert the twistor representatives for momentum eigenstates:
$$
\varphi_i (\omega,\pi)=\frac{1}{2}\int \D u_i \;\rd k_i \;k_i \; \delta^4(\pi_A-k_i \lambda_{i  A})\; \e^{-\frac{\ii}{k}\widehat{\lambda}_{i  A}\omega^A}
$$
where a factor of one half has been introduced for later convenience. 

The scattering amplitude is then
\begin{multline*}
{\cal A}(P_1,P_2,P_3)=\frac{1}{8}\int \D^3\pi \;\rd^4\omega\;\delta(\omega\cdot \pi)\left(\prod_{i =1}^3\int\D u_{i }\;\rd k_{i }\;k_{i }\;\delta^4(\pi_A-k_{i }\lambda_{i  A})\right)\; \\
\times \exp\left(-\ii\sum_{i =1}^3\frac{\omega^A\widehat{\lambda}_{i  A}}{k_{i }}\right) \, .
\end{multline*}
 Writing the quadric delta-function as an integral
$$
\delta(\omega\cdot\pi)=\int\rd t\;\e^{\ii \,t\,\omega\cdot\pi} \, ,
$$
 then doing the four $\omega^A$ integrals, gives
$$
{\cal A}(P_1,P_2,P_3)=\frac{1}{8}\int \D^3\pi\;\rd t\; \left(\prod_{i =1}^3\int\D u_{i }\;\rd k_{i }\;k_{i }\;\delta^4(\pi_A-k_{i }\lambda_{i  A})\right)\;\delta^4\left(t\,\pi_A-\sum_{i =1}^3\frac{\widehat{\lambda}_{i  A}}{k_{i }}\right) \, .
$$
We can already see
that this expression has the support of the 6-momentum conserving delta-functions.
Consider the skew product of the argument of the delta-function $\delta^4\left(t\pi_A-\sum_{i }\frac{1}{k_{i }}\widehat{\lambda}_{i A}\right)$  
with  $\pi$
$$
t\pi_{[A}\pi_{B]}-\sum_{i }\frac{1}{k_{i }}\pi_{[A|}\widehat{\lambda}_{i  |B]}=0 \, .
$$
The first term vanishes and on the support of the $\delta^4(\pi_A-k_{i }\lambda_{i  A})$, the second term may be written as
$$
0=\sum_{i }\lambda_{i [A|}\widehat{\lambda}_{i  |B]}=\sum_{i }P^{i }_{AB} \, ,
$$
which gives the expected momentum conservation. Thus
\begin{equation}\label{3amp}
{\cal A}(P_1,P_2,P_3)={\cal K}\delta^6\left(\sum_{i =1}^3P^{(i )}_{AB}\right) \;.
\end{equation}
for some ${\cal K}$.  Since there are no Lorentz invariants of three
null momenta that add up to zero, we see that ${\cal K}$ must be
constant and so we have the correct amplitude.  A more laborious
argument can be used to obtain ${\cal K}$ explicitly.

\section{Discussion: Conformal Theories in Twistor Space}

One of the triumphs of the twistor programme was the elegant
description of self-dual Yang-Mills in four dimensions in terms of the
Penrose-Ward correspondence \cite{Ward:1977ta}.  This correspondence
relates a holomorphic vector bundle on projective twistor space
without connection to a holomorphic bundle over space-time with
self-dual connection. In six dimensions, a connection on a fibre
bundle cannot be self-dual; however, a gerbe can have self-dual
connection. In this section we shall be particularly interested in
six-dimensional linearised physical theories in which self-dual gerbes
play a role. These objects play an important role in string theory and
it is doubtful that a full understanding of M-theory will be possible
without at least a partial understanding of the conjectured non-linear
versions of such theories.

\subsection{Self-dual Gerbes in Twistor Space}
An abelian gerbe on space-time is usually thought of as a
generalization of a connection in which the connection 1-form is
replaced by a two-form $B$ defined modulo the addition of the exterior
derivative of a 1-form and so the `curvature' $\rd B$ is now a 3-form.   On twistor space, we have the two descriptions: as $b\in H^2(\cO)$ and $g\in H^3(\cO(-6))$ both as forms modulo $\dbar$-exact forms.  Since these are both in a potential modulo gauge format, the form degree of the latter does not naturally fit the concept of the gerbe, but the $H^2$ case does.   
Indeed this case has already been studied as a route to defining a
holomorphic gerbe on twistor space by Hitchin and Chatterjee
\cite{Hitchin:1999fh,Chatterjee} who present the theory as a
generalization of the \v Cech description of line bundles; this is
outlined in appendix \ref{indirect-gerbe} for \v Cech cohomology along with the Penrose transform in the indirect case leading to a potential modulo gauge description for the space-time field.  

The Dolbeault route gives a slightly different picture to that
arising from the \v Cech  approach. In the Dolbeault picture, we can
think of the $b\in H^2(\cO)$ as defining a class of local $\dbar $-operators
$\dbar_{a_i}$ on a fixed complex line bundles over a covering $U_i$ of $Q^I$ for
which $\dbar_{a_i}^2=b$.  In the simplest case these can be
$\dbar$-operators on the same 
trivial line bundle and $b$ is the obstruction to extending the local
$\dbar_{a_i}$ to a global $\dbar$-operator. 

It is straightforward to write down action principles for the linear
theories on twistor space.  
We first define the holomorphic volume form
$\Omega\in\Gamma(Q,\Omega^6(6))$ by 
$$
\int_{Q} (\cdot ) \Omega=\int_{\CP^7}(\cdot ) \; \D^7Z \; \bar\delta^1(Z^2) \, . 
$$
In
Euclidean signature, 
for $\M^I=$ Euclidean $\R^6$, the $S_x$ sweep out $Q^I$.
Then an action for a pair $(g,f)\in \Omega^{0,2}(k-2)\times\Omega^{0,3}(-k-4)$ to define Dolbeault cohomology classes is
$$
S[g,f]=\int_Q g\wedge \dbar f\wedge \Omega \, , 
$$
since solutions to field equations mod gauge give 
$$
([g],[f])\in H^2(\cO(k-2))\times H^3(\cO(-k-4))=\Gamma_k\times
\Gamma_k \, .$$ For $k=2$ this is an action for a pair of self-dual
gerbe fields and for $k=4$ it gives an action for a pair of the
spin-two field discussed below. Such an action for the $k=2$ case 
is echoed on space-time by the action 
$$
S(H,B)=\int_{\M^I} H\wedge \rd B\, , \qquad (H,B)\in \Omega^{3+}\times
\Omega^2\, ,
$$
since self-dual 3-forms annihilate self-dual 3-forms under wedge
product in six dimensions.  Thus it is not so surprising.  

In Lorentz signature we can write action formulations that do not seem
to have a space-time analogue.  Recall that in Lorentz signature we
have the quaternionic reality structure $\pi_A\rightarrow \hat\pi_A $
etc., $Z\rightarrow \hat Z$ with $\hat{\hat Z}=-Z$.  In this
signature, the $S_x$ that are invariant under the quaternionic
conjugation only sweep out the real codimension-one set $$Q_0=\{Z\in
Q|Z\cdot \hat Z=0\} \, .$$ If we now choose data $(g,\tilde g)\in
\Omega^{0,2}(k-2)\times\Omega^{0,2}(-k-4)$ we obtain
$$
S[g,\tilde g]=\int_{Q_0} \tilde g\wedge \dbar g\wedge \Omega \, .
$$
The field equations lead to a pair of cohomology classes but now 
$$
([g],[\tilde g]) \in H^2(Q_0,\cO(k-2))\times  H^2(Q_0,\cO(-k-4))=\Gamma_k\times 0 \, . 
$$
Because of the vanishing of $H^2$s for sufficiently negative homogeneity, we only obtain one  helicity $k/2$ field. Thus an action corresponding to a linear self-dual gerbe theory on Minkowski space is given by
\begin{equation}\label{Mink}
S[b]=\int_{Q_0} \D^7Z\wedge\bar{\delta}(Z^2)\wedge \bar{\delta}(Z\cdot\hat{Z})\wedge \tilde{b}\wedge\bar{\partial}\,b
\end{equation}
for some twistor field $\tilde{b}$ of homogeneity $-4$. The equations
of motion imply that $b\in H^2(Q;\cO)$ as it should and $\tilde{b}\in
H^2(Q;\cO(-4))$; however, $H^2(Q;\cO(-4))=0$ and so $\tilde{b}$ has no
on-shell degrees of freedom and acts simply as a Lagrange multiplier
that vanishes on-shell, constraining $b$ to lie in $H^2(Q;\cO)$.

The scattering amplitude for the scalar theory in split signature
suggests that there is a twistor action for the space-time Lagrangian
(\ref{scallag}), at least in Euclidean and split signature. However,
it is difficult to make this work coherently.  
One is tempted to write
\begin{equation}\label{scat}
S[\varphi,\phi]=\int_{\C\P^7}\D^7 Z\wedge\bar{\delta}(Z^2)\wedge\left(\phi\wedge\bar{\p}\varphi+\frac{1}{6}\,\varphi\wedge \varphi\wedge \varphi\right)
\end{equation}
where $\bar{\delta}(Z^2)$ is a $(0,1)$-form of weight $-2$ and $\D^7 Z=\varepsilon_{\alpha_0 \alpha _1....\alpha _7}Z^{\alpha _0}\p Z^{\alpha _1}\wedge...\wedge\p Z^{\alpha _7}=\D^3\pi\rd^4 \omega$ is the natural projective $(7,0)$ form on $\C\P^7$ of weight $+8$. Here $\varphi$ is a $(0,2)$-form of weight $-2$ and $\phi$ is a $(0,3)$-form of weight $-4$. The form degrees and weights match and so this action makes sense as an action functional. In split signature, the interaction term gives the expression for the three-point amplitude (\ref{3amp}). This action is well-defined on the quadric $Q$ and, under variation with respect to the representatives gives
$$
\bar{\p}\varphi=0 \, , 	\qquad	\bar{\p}\phi+\frac{1}{2}\,\varphi\wedge \varphi=0 \, .
$$
So $\varphi$ corresponds simply to a solution of the wave equation
whereas $\phi$ seems likely to correspond to a solution to an imhomogeneous
wave equation sourced by $\varphi^2$.  

It therefore remains difficult to piece these together to give an
action for the $\Phi^3$ theory on twistor space.  The Lorentzian
formulation seems to be no better as it is not clear how to encode the
cubic interaction as a $(0,5)$-form.

\subsection{Non-geometric gravitational theories}

Up until now we have been concerned with spin-one gerbes, here we extend our considerations to spin-two fields. The on-shell graviton is given by a field strength $\Psi_{AB}^{CD}$ with spinor-helicity polarization data $\Psi^{ab}_{a'b'}$ which has nine degrees of freedom; however, the spin-two field strength $G_{abcd}$ arising from the direct Penrose transform
\begin{equation}\label{G}
G_{ABCD}=\oint_{S_x}\D^3\pi\,\pi_A\pi_B\pi_C\pi_D\;g(\omega,\pi)
\end{equation}
has five on-shell degrees of freedom. Furthermore $G_{abcd}$ is a \emph{chiral} field, whereas the graviton is not. This spin-two field appearing from twistor space is clearly not describing linearised Einstein gravity, but a more exotic six-dimensional relative. It is conjectured that there exists a superconformal $(4,0)$ theory in six-dimensions \cite{Hull:2000zn,Hull:2000rr,Hull:2000ih,Schwarz:2000zg} which includes just such a field; however, we can consider this field in a bosonic context. A discussion of the supersymmetric theory will be presented in \cite{MRT}. The novelty of this theory is that the spin-two field is not a graviton in the conventional sense and is not thought to give rise to a conventional, geometric, theory of gravitation. Rather, the spin-two field is given by a tensor $C_{\mu\nu\lambda\rho}$ with the symmetries of the Riemann tensor and field strength
$$
G_{\mu\nu\lambda\rho\sigma\eta}=3\partial_{\mu}\partial_{[\nu}C_{\lambda\rho]\sigma\eta}+3\partial_{\eta}\partial_{[\nu}C_{\lambda\rho]\mu\sigma}+3\partial_{\sigma}\partial_{[\nu}C_{\lambda\rho]\eta\mu}
$$
which is self-dual
$$
G_{\mu\nu\lambda \rho\sigma\eta}= \frac{1}{3!}\varepsilon_{\rho\sigma\eta}{}^{\kappa \xi \zeta}G_{\mu\nu\lambda \kappa \xi \zeta}  \, .
$$
In terms of spinor notation, the field can be encoded into a potential field $C_{AB}^{CD}$ symmetric in each pair of indices with (linearised) manifestly self-dual field strength
$$
G_{ABCD}=\nabla_{(A|M}\nabla_{|B|N}C_{|CD)}{}^{MN}  \, .
$$
This field, with 5 on-shell `gravi-gerbe' degrees of freedom, is the highest spin member of the $(4,0)$ multiplet. At the linearised level, a dimensional reduction on a circle to five dimensions yields the linearised form of the conventional Einstein maximal supergravity in five dimensions ad it is conjectured that there exists a non-linear $(4,0)$ theory in six-dimensions which gives rise to the full Einstein supergravity in five dimensions \cite{Hull:2000zn}. It is not clear what the full non-linear $(4,0)$ theory should look like but it is expected that the interactions will not be of a conventional field-theoretic type but rather should be based on yet to be identified M-theoretic principles. In this section we consider only the linearised for of the $(4,0)$ theory in supertwistor space. The spin two field may be described in terms of the conventional Penrose transform (\ref{G}) where $g(\omega,\pi)\in H^3(Q^I;{\cal O}(-8))$ and it is straightforward to generalise the arguments above for the spin-one gerbe to a representative of $H^2(Q^I;{\cal O}(+2))$ to get a description of this field in terms of a potential $C_{ABCD}$, modulo gauge-invariance, from the indirect Penrose transform.

In a subsequent paper we will turn to the supersymmetric formulations, nonconformally invariant and non-chiral theories and reductions to lower dimensions.

\begin{center}
\textbf{Acknowledgements}
\end{center}
We would like to thank Mike Eastwood, Rod Gover, Edward Witten and Martin Wolf for useful discussions.

One of the authors (ATC) is funded by a SoMoPro (South Moravian Programme) Fellowship: it has received a financial contribution from the European
Union within the Seventh Framework Programme (FP/2007-2013) under
Grant Agreement No. 229603, and is also co-financed by the South
Moravian Region.
Both LM and RRE were supported by the EPSRC grant EP/F016654/1.
LM is supported by a Leverhulme Research Fellowship.

\begin{appendix}
\section{The Indirect Penrose transform for a self-dual
  gerbe}\label{indirect-gerbe} 
It is perhaps simplest to understand what a holomorphic gerbe is by comparing its definition with that of a holomorphic line bundle \cite{Hitchin:1999fh}. A line bundle may be understood in terms of a set of transition functions between open sets $g_{ij}:U_i\cap U_j\rightarrow S^1$ with $g_{ij}=g^{-1}_{ji}$ and a triviality condition on the overlap of three open sets: $g_{ij}g_{jk}g_{ki}=1$ on $U_i\cap U_j\cap U_k$. The bundle is holomorphic if $g_{ij}$ are holomorphic functions. By contrast a gerbe is defined by functions on a \emph{triple} intersection
$$
b_{ijk}:U_i\cap U_j\cap U_k\rightarrow S^1 \, ,
$$
with $b_{ijk}=b^{-1}_{jik}=b^{-1}_{ikj}=b^{-1}_{kji}$ and the triviality condition on the overlap of four open sets
$$
b_{jkl}b^{-1}_{ikl}b_{ijl}b^{-1}_{ijk}=1 \, ,	\quad \text{on}	\qquad U_i\cap U_j\cap U_k\cap U_l \, .
$$
Crucially a gerbe, unlike a fibre bundle, is \emph{not} a manifold\footnote{An equivalent way to define a gerbe is in terms of transition \emph{line bundles} on $U_i\cap U_j$, as opposed to the transition \emph{functions} that define a line bundle.}. A \emph{holomorphic gerbe} is one for which the $b_{ijk}$ are holomorphic functions.

We can also do differential geometry on gerbes. We can define a connection on a line bundle by $A_i-A_j=g_{ij}^{-1}\rd g_{ij}$ and a field strength, defined over the whole bundle, $F=\rd A_i=\rd A_j$. On a gerbe we may define a connection in a similar way
$$
H=\rd B_i=\rd B_j\;,	\qquad	B_i-B_j=\rd A_{ij}\;,	\qquad	A_{ij}+A_{jk}+A_{ki}=b^{-1}_{ijk}\rd b_{ijk} \, ,
$$
where $B_i$ is the connection\footnote{Following the physics literature, we shall also refer to $B_i$ as `the gerbe'.} and $H$ is a globally-defined closed three-form field strength. A connection on a gerbe is (anti) self-dual if $H=\pm*H$. The six-dimensional analogue of the Penrose-Ward correspondence, which we sketch below,  relates cohomology classes $[b]=\{b_{ijk}\}$ on twistor space to self-dual connections $B$ on space-time.

The simplest twistor representation for such a self-dual gerbe is via the direct Penrose transform given by
$$
H_{AB}(x)=\int_{S_x}\D^3\pi\;\pi_A\pi_B\;h\left(\omega,\pi\right) \, ,
$$
where $h\in \check{H}^3(Q^I;{\cal O}(-6))$ and the integral is taken over the $\C\P^3$ picked out by the incidence relation.   However, this represents the gerbe as a $(0,3)$-form potential modulo gauge which is not appropriate to define a gerbe on twistor space.

The most natural association between twistor cohomology and connections of a gerbe in space-time comes from representatives of $H^2(Q^I,\cO)$ or more geometrically its exponentiation ${H}^2(Q^I;{\cal O}^*)$ given by the indirect Penrose transform.  This defines a holomorphic gerbe on twistor space.  Following \cite{Chatterjee} this can be understood via \v Cech coholomorgy: let $[b]$ be a representative of $\check{H}^2(Q^I;{\cal O}^*)$ and $\{U_i\}$ a Leray cover\footnote{A Leray cover is one for which the open sets have no cohomology so that the \v Cech cohomology agrees with the standard cohomology.} of $Q$. We then have a family of functions of homogeneity degree zero $[b]=\{b_{ijk}\}$ defined on the triple intersection
$$
b_{ijk}:U_i\cap U_j\cap U_k\rightarrow \C^* \, , \quad b_{ijk} b_{jkl}b_{kli}b_{lij}=1.
$$
Here $b_{ijk}$ is a cohomology representative and is defined on $Q$ and may be lifted to the correspondence space $\F=\{(x,[\pi])\in\M\times \C\P^3\}$ where, by virtue of being pulled back from twistor space, it satisfies
$$
\mu^*\left(\pi_B\nabla^{AB}b_{ijk}\right)=0 \, .
$$
Restricting to $S_x=\C\P^3$, $\check{H}^2(S_x;{\cal O})=0$ and so we can write
\begin{equation}\label{split1}
b_{ijk}=a_{ij}a_{jk}a_{ki} \, .
\end{equation}
Here $a_{ij}=a_{ij}(x,\pi)$ is not pulled back from twistor space (assuming $[b]$ was not trivial) and so $\pi_B\nabla^{AB}a_{ij}\neq 0$. 
We can define $a_{ij}{}^A=\pi_B\nabla^{AB}\log a_{ij}$, and differentiating \eqref{split1} we obtain
$$
a_{ij}{}^A+a_{jk}{}^A+a_{ki}{}^A=0 \, .
$$
It is also the case that $\check{H}^1(S_x;{\cal O})=0$ and so 
\begin{equation}\label{split2}
a_{ij}{}^A=f_i{}^A-f_j{}^A\, .
\end{equation}
Furthermore, since partial derivatives commute,  $\pi_C\nabla^{C[A}a_{ij}^{B]}=0$. Thus differentiating \eqref{split2} we obtain
$$
\pi_C\nabla^{C[A}f^{B]}_i=\pi_C\nabla^{C[A}f^{B]}_j=S^{AB}
$$
The left and right hand sides of the first equality are defined on different patches ($U_i$ and $U_j$) but are equal. From this we infer the existence of the globally defined field $s^{AB}$ which is homogenous of degree two in $\pi_A$.  We can therefore  express the $\pi$ dependence  explicitly as $s^{AB}=s^{ABCD}\pi_C\pi_D$. We note that skew symmetry implies that $\pi_A\pi_B\nabla^{AB}=0$ so that $\pi_{A}a^A_{ij} =0= \pi_Af^A_{ij}$ and $\pi_AS^{AB}=0$.  Thus $s^{ABCD}$can be expressed in terms of some $B_A^B$ by
$$
s^{ABCD}=\frac{1}{2}\varepsilon^{ABE(C}B_E^{D)} \, .
$$ 
This potential can be taken to be traceless, corresponding to a 2-form, and is defined modulo gauge 
$$
B_A^B\sim B_A^B+\nabla_{AC}A^{BC}-\frac14 \delta^B_A \nabla\cdot A
$$ 
for a 1-form $A^{BC}=-A^{CB}$ because $f_i^A$ was defined up to $f_i^A\rightarrow f_i^A+ \varepsilon^{ABCD}\pi_BA_{CD}$. $B_A^B$ can therefore be interpreted as a gerbe connection on space-time. 
We have that $\pi_A\nabla^{A[B}S^{CD]}=0$ from its definition, and this gives the field equation
$$
\nabla^{A(B}B_A^{C)}=0\, .
$$
The gauge-invariant field strength for the gerbe is therefore
$$
H_{AB}=\nabla_{(A|C}B_{|B)}^C \, ,
$$
corresponding to a self-dual three-form which is closed, being the exterior derivative of the two-form corresponding to $B$.

\end{appendix}

\end{document}